\documentclass[iop]{emulateapj}

\usepackage{aas_macros}

\usepackage{epsfig}
\usepackage{longtable}
\usepackage[breaklinks]{hyperref}
\usepackage{multirow}

\makeatletter
\def\LT@makecaption#1#2#3{%
  \LT@mcol\LT@cols c{\hbox to\z@{\hss\parbox[t]\LTcapwidth{%
    \sbox\@tempboxa{#1{#2 }#3}%
    \ifdim\wd\@tempboxa>\hsize
      #1{#2 }#3%
    \else
      \hbox to\hsize{\hfil\box\@tempboxa\hfil}%
    \fi
    \endgraf\vskip\baselineskip}%
  \hss}}}
\makeatother

\def\bfref{}

\begin{document}

\title{Pulsar Observations Using the First Station of the Long Wavelength Array and the LWA Pulsar Data Archive}

\author{K.~Stovall$^{1}$, P.~S.~Ray$^{2}$, J.~Blythe$^{2}$, J.~Dowell$^{1}$, T.~Eftekhari$^{1}$, A.~Garcia$^{3}$,
T.~J.~W.~Lazio$^{4}$, M.~McCrackan$^{1}$, F.~K.~Schinzel$^{1}$, G.~B.~Taylor$^{1}$ }


\footnotetext[1]{Department of Physics and Astronomy, University of New Mexico, Albuquerque, NM, USA; stovall.kevin@gmail.com}
\footnotetext[2]{Space Science Division, Naval Research Laboratory, Washington, DC 20375-5352, USA}
\footnotetext[3]{Center for Advanced Radio Astronomy, University of Texas at Brownsville, One West University
Boulevard, Brownsville, Texas 78520, USA}
\footnotetext[4]{Jet Propulsion Laboratory, California Institute of Technology, 4800 Oak Grove Drive, Pasadena, CA, 91106, USA}

\begin{abstract}
We present initial pulsar results from the first station of the Long Wavelength Array
(LWA1) obtained during the commissioning period of LWA1 and early science results. We
present detections of periodic emission from {\bfref 44} previously known pulsars,
including 3 millisecond pulsars (MSPs). The effects of the interstellar medium on pulsar
emission are significantly enhanced at the low frequencies of the LWA1 band (10--88 MHz),
making LWA1 a very sensitive instrument for characterizing changes in dispersion measures
(DM) and other effects from the interstellar medium. {\bfref Pulsars also often have
significant evolution in their pulse profile at low frequency and a break in their spectral
index.} We report DM measurements
for {\bfref 44} pulsars, mean flux density measurements for {\bfref 36} pulsars, {\bfref
and multi-frequency component spacing and widths for 15 pulsars with more than one profile component}.
{\bfref For 27 pulsars, we report spectral index measurements within our frequency range.}
We also introduce the LWA1 Pulsar Data Archive, which stores reduced data products from
LWA1 pulsar observations. Reduced data products for the observations presented here can be
found on the archive. Reduced data products from future LWA1 pulsar observations will also
be made available through the archive.
\end{abstract}
\maketitle

\section{Introduction}\label{sec:intro}
Though pulsars were originally discovered at 81.5 MHz~\citep{1968Natur.217..709H} and much
of the initial follow-up of them were conducted at low frequencies (below 200 MHz), the
majority of pulsar observations conducted after the mid-1970s moved to frequencies
of about 350 MHz and above. This move to higher frequencies {\bfref was} largely due to three reasons.
The first is the detectability of pulsars due to {\bfref radio wave propagation effects through}
the interstellar
medium (ISM), in particular dispersion and scattering. The time delay as a function of
frequency ($\nu$) due to the dispersive properties of the ISM {\bfref scales} as $\nu^{-2}$. Though
this effect is largely correctable through the use of incoherent dedispersive techniques and
completely correctable using coherent dedispersive techniques, these techniques require
large computational resources and have only recently begun to be viable for {\bfref regular use
in} low frequency
observations. The effects of interstellar scattering broaden a pulse roughly as $\nu^{-4}$.
These effects are not correctable without a clear detection of a scattered pulse and therefore
a pulse which is completely scattered out cannot be recovered. The second reason which led
to the majority of pulsar observations being conducted at higher frequencies is that many
pulsars have an intrinsic spectral turnover at around 100-200 MHz. Therefore, observations
at higher frequencies often result in higher quality detections.  A third reason pulsar
observations moved to higher frequency is the large sky temperature from Galactic synchrotron
emission. Pulsars are generally steep spectrum sources with typical spectral indices {\bfref for
their flux densities} of about {\bfref $-$1.4~\citep{2013MNRAS.431.1352B}}. However, the Galactic
background is {\bfref also bright at low frequencies and has a brightness temperature} spectral
index of $-$2.6~\citep{1982A&AS...47....1H}.

\subsection{Low Frequency Profile Evolution and Spectral Turnover}
Despite the previously described difficulties in observing pulsars at low frequencies, such
observations are necessary to fully understand the pulsar emission mechanism. Many effects
that are intrinsic to the pulsar are observed, such as {\bfref intrinsic} profile evolution
and {\bfref a} spectral turnover {\bfref in a large fraction of pulsars}.
{\bfref A typical form of intrinsic profile evolution is the so-called ``radius-to-frequency''
(RFM) mapping effect, which postulates that lower frequencies are emitted
higher in the pulsar's magnetosphere, resulting in wider pulse profiles at low frequency
than at high frequency~\citep{1978ApJ...222.1006C}. Observationally, RFM is most prominent at
the lowest frequencies and may only be exhibited below about 1~GHz~\citep{1991ApJ...377..263T}.
In \cite{1983ApJ...274..333R} and other papers in a series on the pulsar emission mechanism,
the authors modeled pulsar emission as `core' emission coming from near the magnetic pole
and `cone' emission from a number of concentric cones. The authors noted that `core' emission
tends to have a steeper spectral index versus that of `cone' emission and theorized that the
two may even be the result of different emission mechanisms. Another model put forward
by~\cite{1988MNRAS.234..477L} explains the emission from pulsars as randomly located patches
of emission within a `window function.' Other models which combine aspects of the above two models
have also been suggested. For example,~\cite{2007MNRAS.380.1678K} postulated that patchy
emission for a particular frequency could originate at varying heights within the pulsar's magnetosphere.
In this model, the `core' emission is coming from closer to the neutron star's surface while
the `cone' emission comes from higher in the magnetosphere. Measurements of profile evolution
for a wide range of frequencies for a large population of pulsars are necessary to constrain
these theories.

At frequencies above a few hundred MHz, the pulsar spectrum is typically modeled as a power
law of the form $S_{\nu} \propto \nu^{-\alpha}$, where typical values for $\alpha$ are around
{\bfref $-$1.4~\citep{2013MNRAS.431.1352B}}.
However, many pulsars exhibit a low frequency turnover in their spectrum in the 100-200 MHz
range~\citep[see for example,][]{1994A&A...285..201M}, others have a spectral turnover at higher
($\sim$1 GHz) frequencies~\citep[e.g.][]{2007A&A...462..699K,2014MNRAS.445.3105D}, and still others
have not shown a turnover at any frequency at which they have been detected~\citep{1994A&A...285..201M}.
} Development of a complete theory for the pulsar emission mechanism must account for {\bfref the
observed profile evolution described above and these spectral properties}, therefore further
characterization of these effects in a large number of pulsars is warranted.

\subsection{Dispersion and Scattering by the ISM}
{\bfref For an overview of the effects of the ISM on pulsar signals, see~\cite{2002ASPC..278..227C}.
Two of the ISM effects are dispersion and scattering, whose effects follow $\nu^{-2}$ and $\nu^{-4}$ dependencies,
respectively. Due to these dependencies, these effects are significantly stronger and can be very precisely measured
at lower frequencies and, therefore measurements of the dispersion measure (DM) and scattering can be
used to monitor the characteristics of the ISM along the lines-of-sight to
pulsars. Such precise measurements enable the testing of the cold plasma dispersion model~\citep{1991ApJ...373L..63P} and
deviations from the $\nu^{-4}$ dependence of pulse broadening~\citep{2001ApJ...562L.157L}.

Another potential application of these precise measurements are to correct for time dependent
dispersion and scattering changes in high precision pulsar timing such as is done in pulsar timing
array (PTA) experiments attempting to detect low frequency gravitational wave
emission~\citep{2011MNRAS.414.3117V,2012arXiv1210.6130M,2013ApJ...762...94D}. The efforts to detect
gravitational waves (GWs) using PTAs are greatly constrained by the ability to understand and mitigate
various sources of noise in the MSP timing residuals. Important sources of noise in the timing
residuals are DM variations in time and frequency and pulse broadening due to scattering~\citep{2010arXiv1010.3785C}.
It is currently unclear whether or not observations of DM and scattering effects at low frequencies
can be used to directly correct PTA datasets. Simulations suggest that measurements of DM variations at
low frequencies are unlikely to be useful in direct correction for DM variation~\citep{2015arXiv150308491C}.
However, such observations will at the very least provide a better understanding of the ISM and could prove
to at least partially correct for these variations.
}

\subsection{New Low Frequency Facilities}
In recent years, many low-frequency observatories have come on line, including the first
station of the  Long Wavelength Array~\citep[LWA1; described in more detail below, but also see][]{2012JAI.....150004T},
the Low Frequency Array~\citep[LOFAR;][]{2013A&A...556A...2V,2011A&A...530A..80S}, and the
Murchison Widefield Array~\citep[MWA;][]{2013PASA...30....7T,2015PASA...32....5T}. The development of these
systems is largely due to interest in detecting the epoch of re-ionization. While the
detection of signatures from the epoch of reionization was a key requirements driver for
LOFAR and MWA, LWA1 originated with the goal of creating a multi-purpose instrument operating
in a relatively
unexplored region of the radio spectrum. {\bfref Another low frequency instrument, the
Ukrainian T-shaped Radio telescope, second modification~\citep[UTR-2;][]{1978Ap&SS..54....3B}
has recently received an upgraded backend system~\citep{2010A&A...510A..16R}, making it
significantly more sensitive to radio pulsar emission~\citep{2013MNRAS.431.3624Z}. }
With these {\bfref new systems} beginning to come on line
and with {\bfref computational resources now capable of mitigating} the effects of dispersive smearing,
observations of pulsars at frequencies below about 200 MHz are beginning to be performed
more regularly. {\bfref Pulsar detections for 40 pulsars at about 20 MHz using UTR-2 were
reported in~\cite{2013MNRAS.431.3624Z}, 100 pulsars at about 150 MHz as well as 27 pulsars
at 40 MHz have been reported in~\cite{2015Pilia}, and this work reports on detections of
44 pulsars in the 30 to 88 MHz range. In addition to detections from a large number of
pulsars, studies of individual pulsars have also begun to be performed.
Examples include observations of MSPs with LWA1 and the
MWA~\citep{2013ApJ...775L..28D,2014ApJ...791L..32B}, observations of the mode-switching
PSR B0943+10~\citep{2014A&A...572A..52B}, and a study of the effects of profile evolution
on measured DMs~\citep{2012A&A...543A..66H}.
}

In this paper, we present pulsar observations made with LWA1 and introduce
the LWA Pulsar Data Archive. In
Section~\ref{sec:lwa1}, we will describe LWA1. In Section~\ref{sec:PDP}, we will present
how LWA1 pulsar data has been processed. In Section~\ref{sec:PDA}, we will introduce the
LWA1 Pulsar Data Archive and how to access pulsar data from previous observations. In
Section~\ref{sec:dmandflux} we will describe the methods we used to measure the DM and
mean flux density. In Section~\ref{sec:results}, we will detail current observations of known pulsars using
LWA1.

\section{The First Station of the Long Wavelength Array}\label{sec:lwa1}
Here we briefly describe LWA1 as relevant to pulsar observations, for a detailed, general
description of LWA1 see~\cite{2012JAI.....150004T} and~\cite{2013ITAP...61.2540E}. 
LWA1 is capable of tracking four sky locations using independent delay-and-sum beams. These
beams each have two independent frequency ranges called tunings, each with dual polarization.
{\bfref These tunings} can have a center frequency in the range of 10-88 MHz {\bfref and have} a frequency tagging
better than 1 mHz~\citep{memo201}. The LWA1 data are timetagged by the digital processor using a system clock-based counter
that is synchronized with the station's GPS receiver.  The GPS receiver provides both the
absolute time, as well as the signals needed to generate the system clock: a 1 pulse per
second and a 10 MHz signal.  The 10 MHz signal is generated using an internal rubidium
oscillator that provides an accuracy better than one part in 10$^{11}$ when the GPS
receiver is locked. LWA1 can be operated in full-bandwidth mode which does not filter any of
the 10-88 MHz frequency range, or in
split-bandwidth, which attenuates signals received below about 30 MHz. A majority of pulsar
observations have been taken in split-bandwidth mode in order to {\bfref mitigate} the effects of RFI
at frequencies below about 25 MHz. Typically, pulsar observations are recorded in baseband
mode at a sample rate of 19.6 megasamples per second. {\bfref Since the samples are complex, this allows}
a bandwidth {\bfref up to} 19.6 MHz ({\bfref 17-18}
MHz usable). In cases where the data rate for this mode is too large and the {\bfref desired} 
science can be accomplished with less bandwidth, data can be recorded using lower sample rates.
{\bfref Results presented here (except where otherwise noted) were taken in split bandwidth
mode with the full 19.6 MHz of bandwidth in baseband mode and then were converted into a filterbank
with 4,096 channels and a sample time of 209 $\mu s$ using tools described in
Sec.~\ref{sec:PDP} and~\cite{2012JAI.....150006D}.
}

{\bfref 
The RFI environment of LWA1 varies with time of day and observing frequency. RFI is
generally stronger during the day, especially at the lower part of the LWA1 band (below about
25 MHz). LWA1 was designed to operate between where the Earth's Ionosphere becomes transparent
to radio waves at about 10 MHz up to the lower part of the FM band at 88 MHz. There are some
strong frequency carriers within the LWA1 band (for example, analog TV channel 2 at 55.25 MHz),
but most of the band is generally clean. Other sources of terrestrial RFI in the LWA1 band are
CB and HF radio transmissions, but they are intermittent and dependent on the state of the
ionosphere and therefore are not a problem much of the time. Pulsar observations presented in Section~\ref{sec:results}
were processed using a RFI mask that we calculated
using the \texttt{rfifind} tool from the Pulsar Search and Exploration Toolkit\footnote{http://www.cv.nrao.edu/\textasciitilde sransom/presto/}~\citep[\texttt{PRESTO};][]{2001PhDT.......123R} pulsar
reduction package. Table~\ref{table:RFI} gives the fraction of observations with percentages
of data masked below 5\%, 10\%, 25\%, and 50\% for various frequencies and times of day.
Frequencies near the edge of the LWA1 band are more affected by RFI than the middle section
of the band. Since most of our observations were done above 30 MHz, we do not see much variation
in RFI as a function of time of day, though the cleanest time is from 0 to 4 hours local time.

\setlength{\tabcolsep}{3pt}

\begin{table*}
\begin{center} {\footnotesize
\begin{tabular}{cccccc|cccccccc}
\tablecaption{RFI Masked Fractions for LWA1}
\label{table:RFI}
Freq (MHz) & N & $<$5\% & $<$10\% & $<$25\% & $<$50\% & Hour (local time) & N & $<$5\% & $<$10\% & $<$25\% & $<$50\% \\
\hline\hline
30 - 40 & 68 & 0.68 & 0.78 & 0.84 & 0.93 & 00 - 04 & 42 & 0.79 & 0.93 & 1.00 & 1.00 \\
40 - 50 & 64 & 0.78 & 0.89 & 0.98 & 1.00 & 04 - 08 & 54 & 0.76 & 0.93 & 0.94 & 0.98 \\
50 - 60 & 44 & 0.84 & 0.91 & 1.00 & 1.00 & 08 - 12 & 23 & 0.78 & 0.78 & 0.91 & 0.96 \\
60 - 70 & 69 & 0.86 & 0.90 & 0.97 & 0.99 & 12 - 16 & 61 & 0.72 & 0.82 & 0.90 & 0.95 \\
70 - 80 & 94 & 0.64 & 0.86 & 0.95 & 0.96 & 16 - 20 & 78 & 0.76 & 0.94 & 0.97 & 0.99 \\
 & & & & & & 20 - 24 & 99 & 0.68 & 0.76 & 0.91 & 0.94 \\
\hline
\end{tabular}
\caption[RFI Masked Fractions]{{\bfref Left: Fraction of datasets within various frequency ranges with percentages of data masked below 5\%, 10\%, 25\%, and 50\%. Right: Fraction of datasets with percentages of data masked below 5\%, 10\%, 25\%, and 50\% as a function of local time.} } }
\end{center}
\end{table*}

}

\section{LWA1 Pulsar Data Processing}\label{sec:PDP}
A set of software tools available for processing data from LWA1 are available in the LWA
Software Library\footnote{\url{http://fornax.phys.unm.edu/lwa/trac/wiki}}~\citep[LSL;][]{2012JAI.....150006D}.
A number of tools useful for processing pulsar data are included in the `Pulsar' extension
to LSL. A majority of pulsar observations are processed with \texttt{writePsrfits2.py}, a
tool that synthesizes a filterbank from the raw beam data and stores the results in
PSRFITS~\citep{2004PASA...21..302H} format. The data can then be further processed using
typical pulsar data reduction methods.

Due to the extreme effects of dispersion at LWA1 frequencies, many pulsars require coherent
dedispersion be applied in order to avoid smearing of the pulses within a frequency channel.
For these pulsars, the
\texttt{writePsrfits2D.py} can be used. This tool is similar to \texttt{writePsrfits2.py}, but
coherent de-dispersion is applied on a per-channel basis after the filterbank.  The
de-dispersion method is an optimized C-based implementation of the de-dispersion module
available in LSL.  The C-based code uses both OpenMP to parallelize the process across
channels and the FFTW library~\citep{Frigo98fftw:an} for optimized Fourier transforms.  The placement of
the de-dispersion after the filterbank allows for this process to be more memory efficient
since the dispersion delay within a single channel is less than the sub-integration block size.

\begin{figure}[h!]
\centering
\includegraphics[width=0.35\textwidth,angle=270]{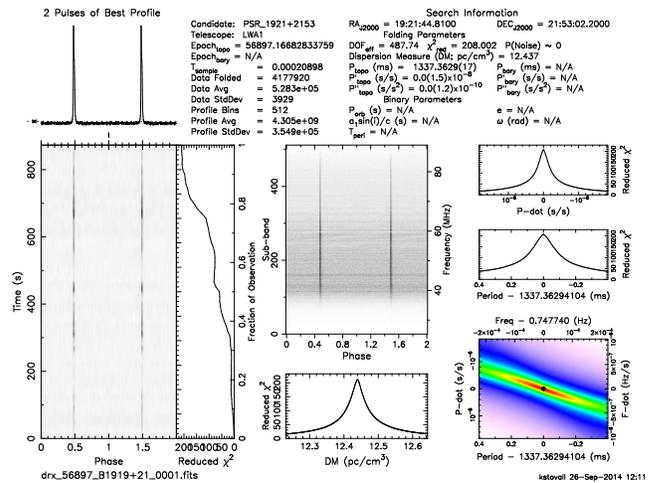}
\caption[B1919+21]{PRESTO diagnostic plot of a 15 minute integration on PSR B1919+21 covering the entire
split-bandwidth configuration of LWA1 using 2 LWA1 beams{\bfref, that has been combined using \texttt{LSL}'s \texttt{writePsrfits2Multi.py} tool}. }
\label{fig:wholebandwidth}
\end{figure}

Though the maximum bandwidth of a single tuning is 19.6 MHz, if the {\bfref center frequency of the} tunings
are {\bfref chosen with appropriate spacing},
they can be combined into a single data file. This includes tunings recorded with a separate beam,
though in this case, the PSRFITS file must be created using \texttt{writePsrfits2Multi.py} or
\texttt{writePsrfits2DMulti.py}, which {\bfref are tools} that aligns the start and end point for files
taken {\bfref simultaneously} using different LWA beams.
Figure~\ref{fig:wholebandwidth} shows a plot of PSR B1919+21 taken with two beams with tuning center frequencies of 35.1,
49.8, 64.5, and 79.2 MHz, each at 19.6 MHz bandwidth. The data were processed using
\texttt{writePsrfits2Multi.py} and then combined in frequency.

\section{LWA Pulsar Data Archive}\label{sec:PDA}
We have begun to store reduced data products from pulsar observations and have made them publicly
available on the LWA Pulsar Data Archive\footnote{\url{http://lda10g.alliance.unm.edu/PulsarArchive/}}.
These data products can be used
for the generation of pulse times-of-arrival (TOAs), analysis of single pulse properties, and analysis of the
pulse profile for each pulsar detection that has been made. Much of the data used in this paper and that {\bfref are}
now available on the archive are the result of commissioning efforts and early pulsar project observations,
however since LWA1 is now operating as an University Radio Observatory and telescope time allocation is
awarded through a proposal system, projects which have been awarded time have a 1-year proprietary period
for their data. Following the proprietary period, reduced data products from pulsar observations will be made available
through the archive. The data products from each observation include the folded result from \texttt{PRESTO}),
folded results in PSRFITS format, folded results from \texttt{DSPSR}\footnote{http://dspsr.sourceforge.net/}, a search
format file with the data
sub-banded using the LWA1 determined DM into 128 frequency channels, a timeseries which has
been de-dispersed at the LWA1 determined DM, a timeseries at DM=0 $\mathrm{pc\;cm^{-3}}$, and RFI masking information. We provide a
more detailed description of the available reduced products below.

\renewcommand{\labelitemi}{$\bullet$}
\begin{itemize}

\item The \texttt{PRESTO} folded results include two {\bfref folded products}; one which uses the best known
ephemeris for the pulsar
and another which starts with the best known ephemeris but searches for the optimal period, period derivative,
and DM for the observation. The results include the folded data, an image of the folded result in postscript
and png formats, and an ASCII text file containing the best profile for that observation. These results are
appropriate for pulse profile analysis and generation of pulsar TOAs.

\item Folded results in PSRFITS format (made using {\bfref \texttt{fold\_psrfits}} and DSPSR) are generated to allow the use of
various tools from \texttt{PSRCHIVE}\footnote{http://psrchive.sourceforge.net/} to analyze the observation. The
data consists of 2048 bins across the profile with 60-s subintegrations, using the full frequency resolution of
the original data file. {\bfref The folded profiles generated using DSPSR are typically made with full Stokes parameters, while
the ones generated using \texttt{fold\_psrfits} contain only total intensity.}

\item The sub-banded data reduces the number of frequency channels to 128 (typically from 4096) by de-dispersing
at the LWA1 determined DM and then summing channels. This data preserves
the individual pulses as well as the original time resolution. The data can be used for generating additional
folded results or for analysis of single pulse characteristics.

\item The individual de-dispersed timeseries loses all of the original frequency information, but preserves the original
time resolution. It can be used for quick folding of data and analysis of individual pulses.

\item The zero-DM timeseries is included so that the RFI environment is known for singlepulse studies as well as to provide
additional information on the general RFI environment of LWA1.

\item RFI mask information, obtained using \texttt{PRESTO}'s \texttt{rfifind} program, has been included to inform data
users of the RFI environment for their particular observation and to provide a long term record of the LWA1 RFI. Typically,
the mask information is obtained by analyzing 10 second long pieces, however this can vary based on the parameters
of the observation. 
\end{itemize}

\section{{\bfref DM, Mean Flux Density, and Pulse Component Analysis}}\label{sec:dmandflux}
For the pulsars that we have detected, we immediately obtain a measurement of their DM and we have estimated their flux density.
For some pulsars, the emission shows the distinct signature of having been scattered by the interstellar medium, but we leave
the measurement of scatter broadening to a future study.

\subsection{DM}\label{subsec:dm}
The measured DM for a pulsar can be time variable, due to motion of the pulsar relative to the Earth and a changing
ISM~\citep[for example;][]{1991ApJ...382L..27P}.
These changes can be stochastic or can have a trend over time. Typically, the DM variations are of the order of 10$^{-3}$
to 10$^{-2}$ $\mathrm{pc\;cm^{-3}\;yr^{-1}}$. Profile evolution and interstellar
scattering can cause systematic errors in the measurement of the absolute DM of a pulsar.
The pulse profile of a pulsar often evolves with
frequency and, therefore, it is not always clear how to properly align template profiles at different frequencies. {\bfref
In~\cite{2007MNRAS.377..677A}, the authors showed through simulation of profile evolution that the true DM can significantly
be affected, particularly for pulsars with complex, asymmetric profiles. \cite{2012A&A...543A..66H} showed
using real pulsar data that this DM error could be corrected by finding a fiducial point in the profile and modeling
the pulse profile as a function of frequency.} Interstellar scattering is also a frequency dependent effect which causes
the centroid of a pulse to be delayed at lower frequencies and can bias the measurement of DM. {\bfref As a starting
point, we used a simplified approach based on single templates (described in more detail below) and leave the more
complex modeling of interstellar scattering and frequency evolution to future work. Due to our more simplified approach,
pulsars with complex profile evolution and interstellar scattering have a bias in their DM measurement, however the error
in these DM measurements are still indicative of the precision with which DMs can be determined with LWA1.}

\begin{figure}[h!]
\centering
\includegraphics[width=0.5\textwidth]{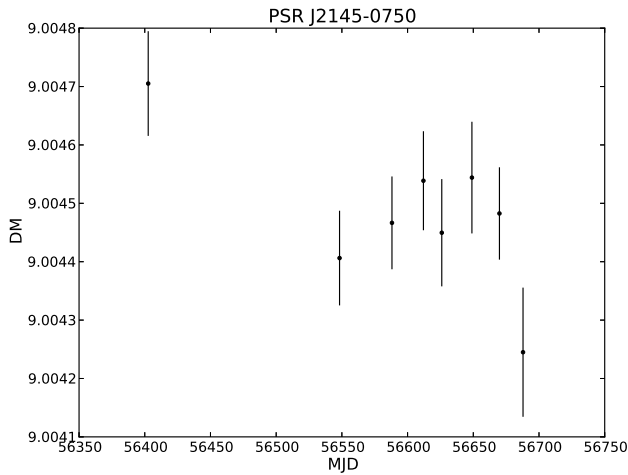}
\caption[J2145-0750 DM]{The DM of PSR J2145-0750 as measured by LWA1 at 8 different epochs.}
\label{fig:J2145DM}
\end{figure}

For all of the pulsars that we detected,
we obtained a measurement of the DM using the following method. {\bfref For slow pulsars, we} folded each of the
observations using ephemerides from the ATNF pulsar catalog. {\bfref The MSPs were folded using ephemerides
from~\cite{2015Demorest} for PSRs J0030$+$0451 and J2145$-$0750 and~\cite{2015Desvignes} for PSR J0034$-$0534.} We then
generated a standard template profile, typically using observations around
50-60 MHz. We then obtained TOAs across the frequency band (typically 4 TOAs per tuning). These TOAs were calculated by
doing a least squares fit in the Fourier domain~\citep{1992PTRSL.341..117T}. We then used
\texttt{TEMPO}\footnote{\url{http://tempo.sourceforge.net}} to fit these TOAs for the best DM for the observation, leaving
all other parameters fixed. The size of errors for our measured DM calculated this way depends
on the pulse width and shape as well as the SNR of the pulsar detection, however typical 1-sigma error values range from about
$10^{-2}\;\mathrm{pc\;cm^{-3}}$ for weak, slow pulsars to $5\times10^{-5}\;\mathrm{pc\;cm^{-3}}$ for millisecond
pulsars. Figure~\ref{fig:J2145DM} shows a plot of the DM for PSR J2145-0750 calculated using the above method for 8
epochs.

\subsection{Mean Flux Density}\label{subsec:flux}
The {\bfref system equivalent flux density (SEFD)} of LWA1 varies with observing frequency,
zenith angle, and {\bfref local sidereal time (LST)}.
In~\cite{memo202} the authors observed the bright known
radio sources, Cyg A, Cas A, Tau A, Vir A, at varying zenith angles and
frequencies. We used the results from this observing program to estimate
the response of LWA1 to varying zenith angle, and used each source to
determine the SEFD {\bfref for Stokes' I} at zenith. The {\bfref SEFD} was determined following the
drift scan method described in~\cite{2013ITAP...61.2540E}. The derived SEFDs
for each object were then {\bfref converted to a SEFD at zenith by} applying
an empirically derived power ratio {\bfref correction} with respect to zenith {\bfref for Stokes' I}: 
\begin{equation}
        \mathrm{SEFD(zenith)}=\mathrm{SEFD(obs)}/\left(166.625*E^{-1.251}+0.401\right),
\end{equation}
where $E$ is the elevation of the observed object in degrees. After this
the {\bfref zenith} SEFD values were averaged per frequency bin. We found that the resulting SEFD is fairly constant across
our observing band, but increases below 40 MHz. We then used the combined measurement of the SEFD
at zenith and the fitted zenith angle dependence to estimate the SEFD for each of our pulsar
observations. The error in the measurement of LWA1's zenith angle SEFD is about 25\%, but to account
for errors caused by SEFD variation with LST (which we are not accounting for in this work) and error
in the fit of the function to zenith angle
dependence, we use a total error of 50\%.

The results from \cite{memo202} are only 
applicable for observations from LWA1 since the last calibration of the LWA1 cable delays, which
occurred on 2013 Feb 28, so we have limited our mean flux density estimates to observations that
occurred after this date. For each observation, we calculated the appropriate SEFD as described
above and used the radiometer equation~\citep{1985ApJ...294L..25D} to estimate the mean flux density.
{\bfref For pulsars with mean flux density measurements at 3 or more frequencies, we then assumed a
power law of the form $S_{\nu}\propto\nu^{\alpha}$ and performed a least squares fit to our measured
mean flux densities.}

{\bfref \subsection{Pulse Component Analysis}\label{subsec:components}
During the Gaussian fitting process, 15 pulsars clearly required multiple Gaussian components,
ignoring pulsars that showed the effects of interstellar scattering. There are 10 and 5 pulsars
requiring 2 components and 3 components, respectively. None of the profiles required more than 3
components. For
these multi-component pulsars, we performed Gaussian fits at each frequency for which we have
data in hand and calculated the widths of each component as a function of frequency as well as
the relative spacing of the components. During this analysis we ordered the components based
on their pulse phase, so that component 1 occurs earliest in the profile. Using the terminology
of~\cite{1983ApJ...274..333R}, the 3-component profiles have conal emission labeled as components
1 and 3, while the bridge or core emission is labeled component 2. At some frequencies
for PSR B0950+08, the phase of the core emission was slightly smaller than the leading edge of
the conal emission. We kept the same component labels and the component spacing for these
cases have a negative sign. For the pulsars with 2 components, we then compared the
relative amplitudes of the fitted Gaussians by calculating the ratio of the two components
$A_{21}=\frac{A_{2}}{A_{1}}$ where $A_{1}$ is the amplitude of the first component (smaller pulse
phase) and $A_{2}$ is the second component (larger pulse phase).
}

\section{Observations \& Results}\label{sec:results}
{\bfref Here we present initial pulsar results using LWA1.} The majority of data presented here were
taken throughout commissioning activities
and during early science runs. In some cases, pulsars were used to verify data integrity for other
science targets. Therefore, they were taken in a variety of observing modes, over a wide range of
dates (2012 May 11 to {\bfref 2015 March 1}), and were not processed in a uniform way.
As of {\bfref 2015 March 1}, LWA1 has been used to detect periodic emission from {\bfref 44} pulsars
and giant bursts from another~\citep[B0531+21;][]{2013ApJ...768..136E}. Large individual pulses have
also been reported from {\bfref PSR} B0950+08~{\bfref\citep{2015AJ....149...65T}}, which is one of
the sources for which we report periodic emission. We present profiles at various frequencies for
{\bfref 44 pulsars} in Fig.~\ref{fig:prof1}. Due to the differences
in data reduction and time difference between observations, the cause of profile offsets from one
frequency to another is not clear. Therefore, for the majority of profiles, we aligned them
manually by shifting the peak of each profile to the middle. However, PSRs B0031-07, B0320+39, and
B0809+74 show considerable profile evolution {\bfref as a function of frequency} throughout the LWA1
frequency band, so these profiles were aligned ``by eye''. Though we leave out an analysis of scattering
due to the ISM, the profile evolution across the LWA1 band for 10 pulsars (PSRs B0329+54, B0450+55, B0823+26, B0919+06,
B1508+55, B1541+09, B1822-09, B1839+56, B1842+14, and B2217+47) show evidence of scattering. We have {\bfref also}
detected the mode switching PSR B0943+10~\citep{SI84,2014A&A...572A..52B} in both bright (B) and
quiet (Q) mode and the resulting profiles are {\bfref shown} in Figure~\ref{fig:prof1}. 

{\bfref \subsection{DM}\label{subsec:resultdm}
Table~\ref{table:detections} contains a list of 44 previously known pulsars detected
through periodicity, the pulsar's period and DM as reported by the ATNF, and our DM measurement
determined as described in Section~\ref{subsec:dm}. {\bfref The two rightmost columns show DM
values reported by~\citep{2013MNRAS.431.3624Z}.} The largest difference in DM between the
ATNF and LWA1 results occurred  for PSR B0031-07, but the change of 0.016 $\mathrm{pc\;cm^{-3}\;yr^{-1}}$
is comparable in magnitude to previously observed DM variation. {\bfref However, as noted in Section~\ref{subsec:dm},
pulsars with asymmetric profiles can have a bias in their DM due to profile evolution. PSR B0031-07 is
seen to have substantial profile evolution in our observations, so this DM bias could account for part
of the DM difference.} Further analysis of DM variations with time will be presented for a sub-sample of
pulsars (including the 3 MSPs J0030+0451, J0034-0534, and J2145-0750) in a future paper.

}

{\bfref \subsection{Mean Flux Density}\label{subsec:resultflux}
We present mean flux densities determined as described in Section~\ref{subsec:flux} {\bfref as well
as the full width at half maximum ($w_{50}$) and full width at 10\% of maximum ($w_{10}$)} for
{\bfref 36} pulsars in Table~\ref{table:flux}. {\bfref We did not measure flux densities for 8 of the pulsars
detected with LWA1, due to only having observations from before 2013 Feb 28, the date of the last cable delay
calibration.
In Figure~\ref{fig:spectra1}, we show our flux density measurements with other reported values at comparable
frequencies~\citep{1973A&A....28..237S,1979SvA....23..179I,1981Ap&SS..78...45I,1998ApJ...509..785S,2011A&A...525A..55K,2013MNRAS.431.3624Z,2014MNRAS.440..327L}.
We have included higher frequency measurements from~\cite{1973A&A....28..237S} to show the general behavior
of the spectrum of each of these pulsars at higher frequency. {\bfref Most of our measurements} are in agreement
with past values, allowing for slight variability in flux and for scintillation. {\bfref Our measurements
indicate larger values for PSRs B0031-07, B0329+54, and B1133+16 than previous measurements. Since our measurements
of flux density are from a single epoch, additional measurements are needed to determine whether these discrepancies
are due to these pulsars being brighter due to favorable scintillation (possibly due to ISM or Ionospheric activity)
or whether it is a systematic effect either of LWA1 or from our SEFD estimation method.} We expect that we will be
able to reduce the size of the error on mean flux measurements in the future by adding additional parameters, {\bfref
such as variation with LST,} to the SEFD determination procedure described in~\ref{subsec:flux}. Even with the
increased flux measurements in the 3 pulsars mentioned above, a general trend of spectral turnover is observed in the
spectra in Figure~\ref{fig:spectra1}.

\begin{figure}[h]
        \includegraphics[trim = 0mm 25mm 0mm 0mm, clip,width=0.5\textwidth]{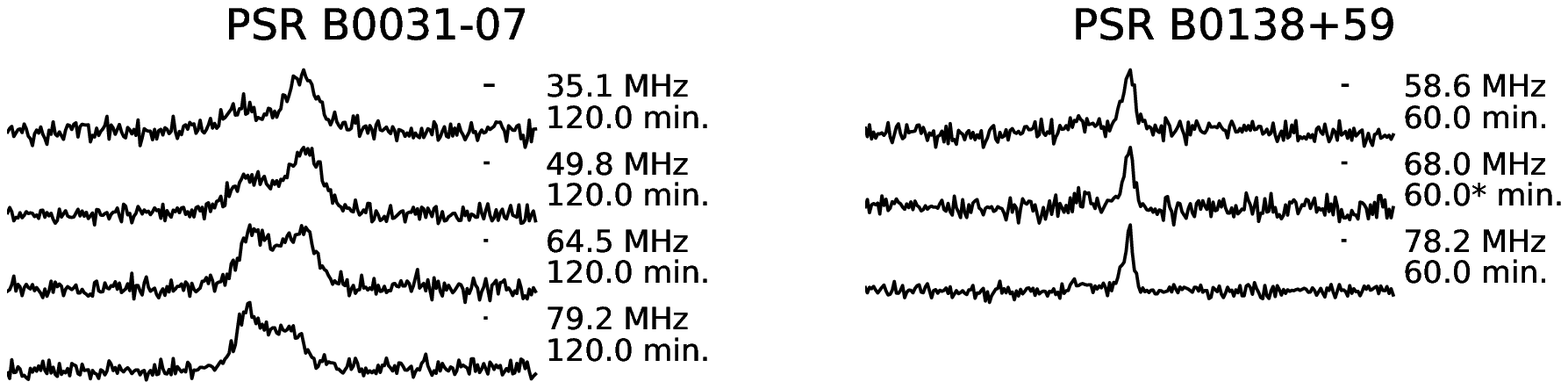}
        \includegraphics[trim = 0mm 25mm 0mm 0mm, clip,width=0.5\textwidth]{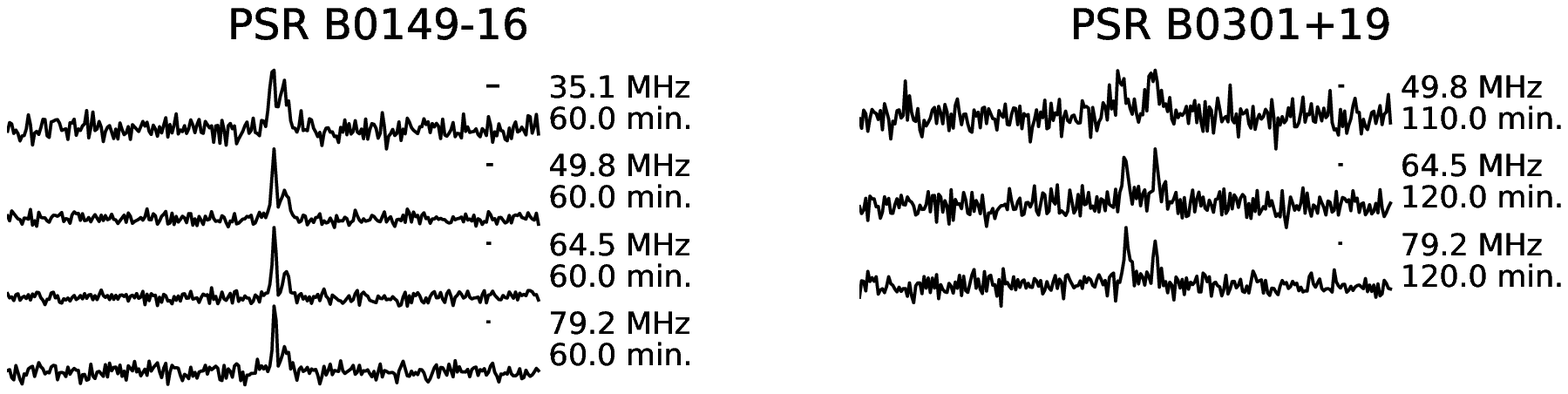}
        \includegraphics[trim = 0mm 25mm 0mm 0mm, clip,width=0.5\textwidth]{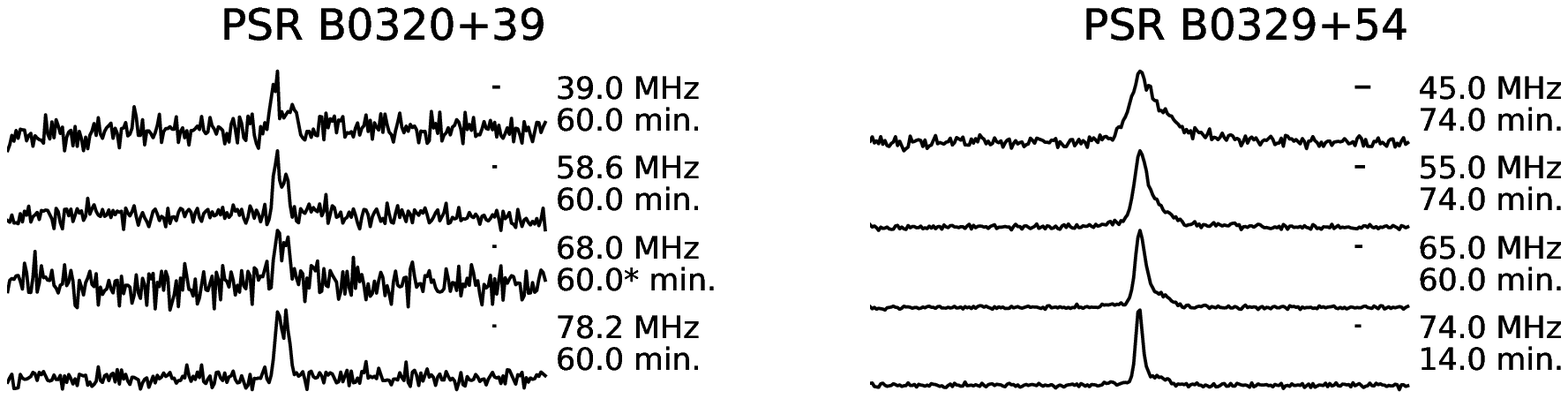}
        \includegraphics[trim = 0mm 25mm 0mm 0mm, clip,width=0.5\textwidth]{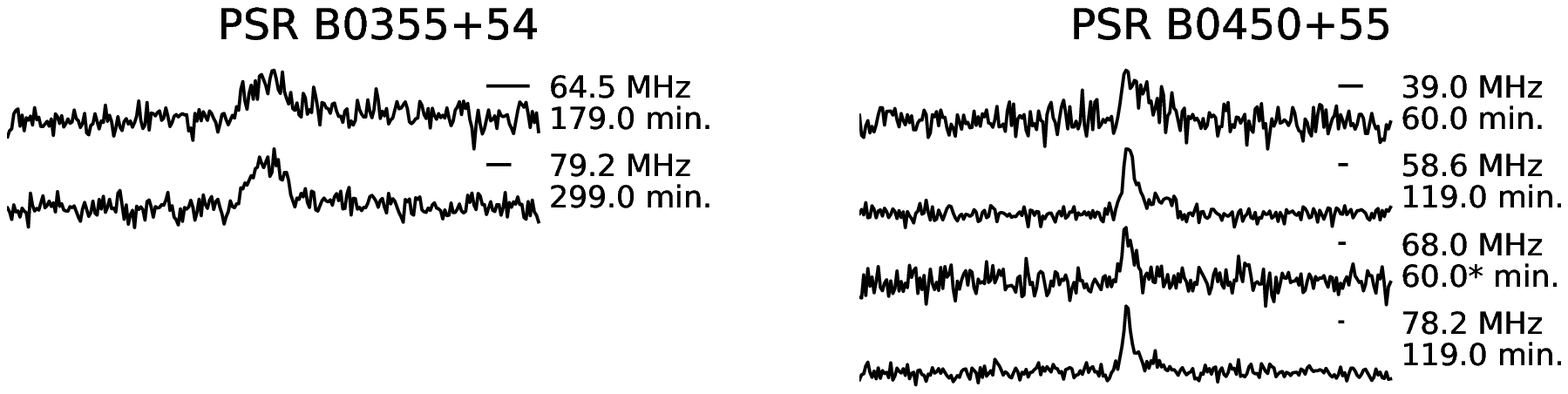}
        \includegraphics[trim = 0mm 25mm 0mm 0mm, clip,width=0.5\textwidth]{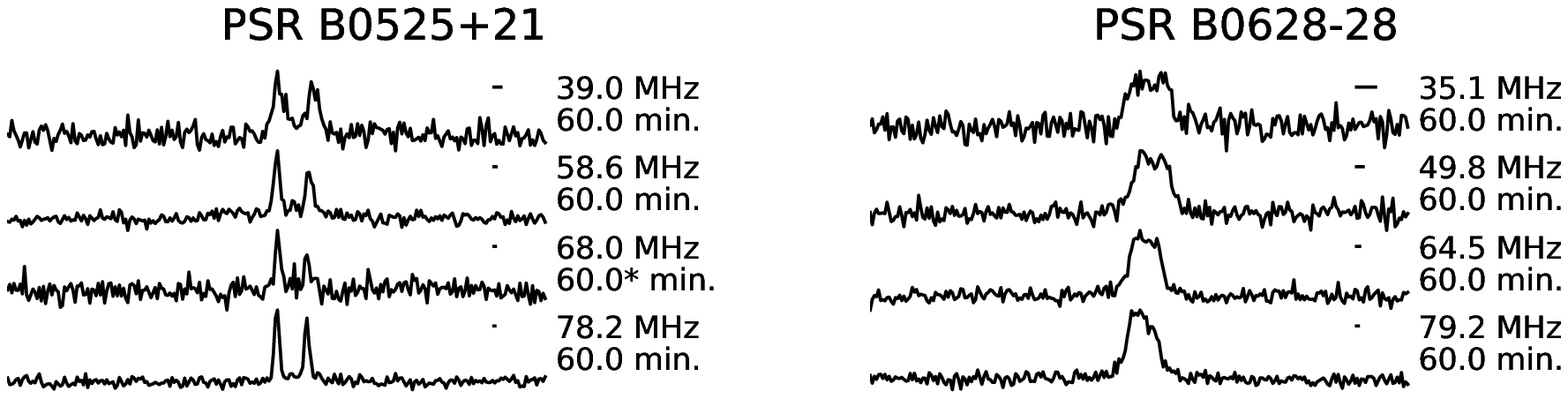}
        \includegraphics[trim = 0mm 25mm 0mm 0mm, clip,width=0.5\textwidth]{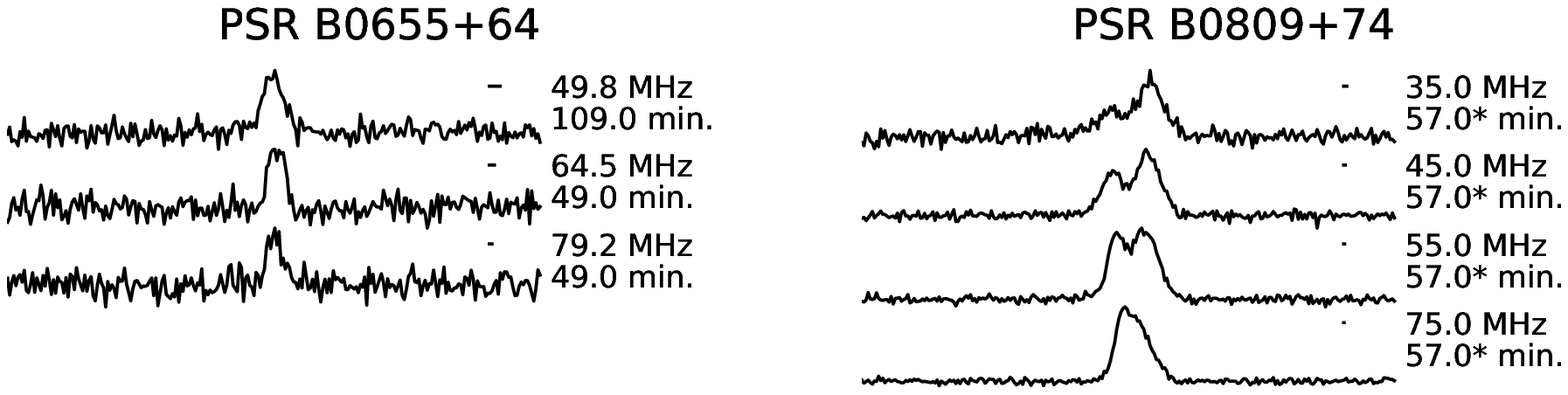}
        \includegraphics[trim = 0mm 45mm 0mm 0mm, clip,width=0.5\textwidth]{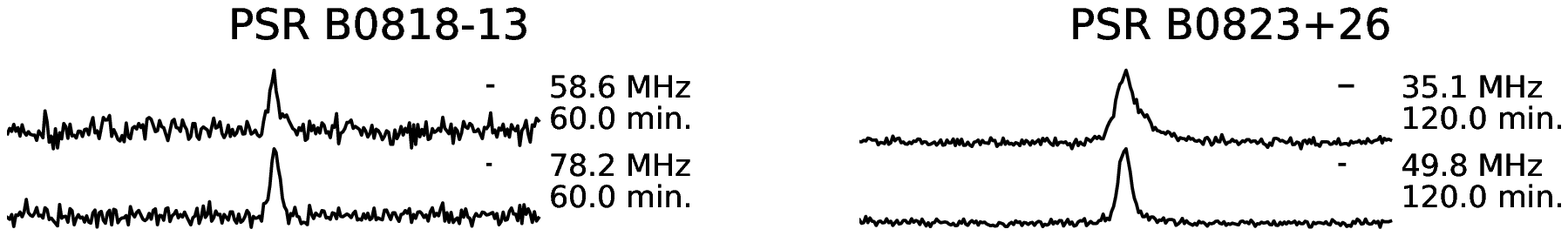}
\end{figure}
\begin{figure}
        \includegraphics[trim = 0mm 25mm 0mm 0mm, clip,width=0.5\textwidth]{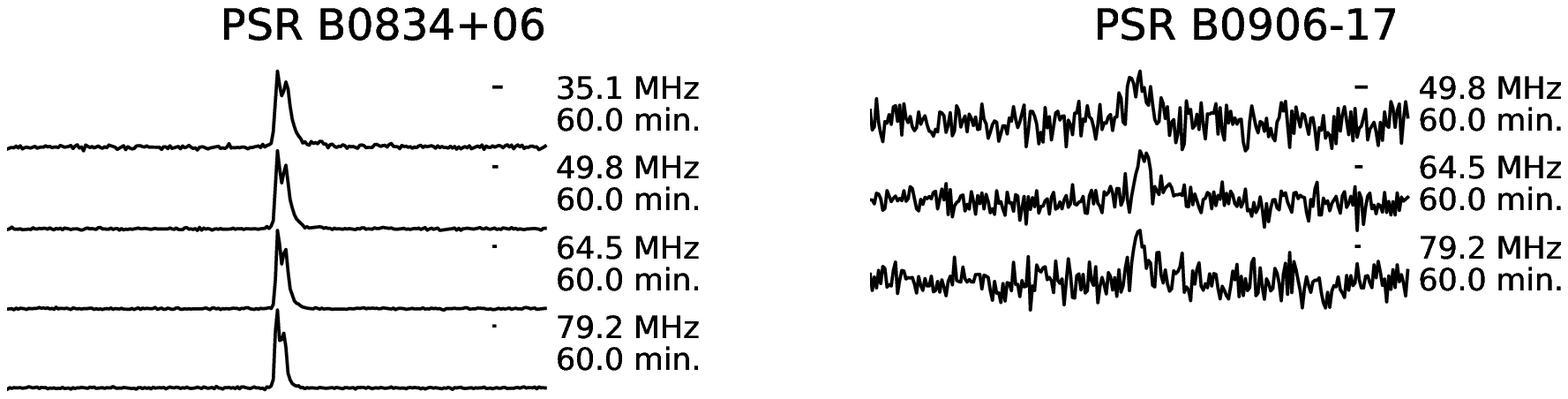}
\end{figure}
\begin{figure}
        \includegraphics[trim = 0mm 15mm 0mm 0mm, clip,width=0.5\textwidth]{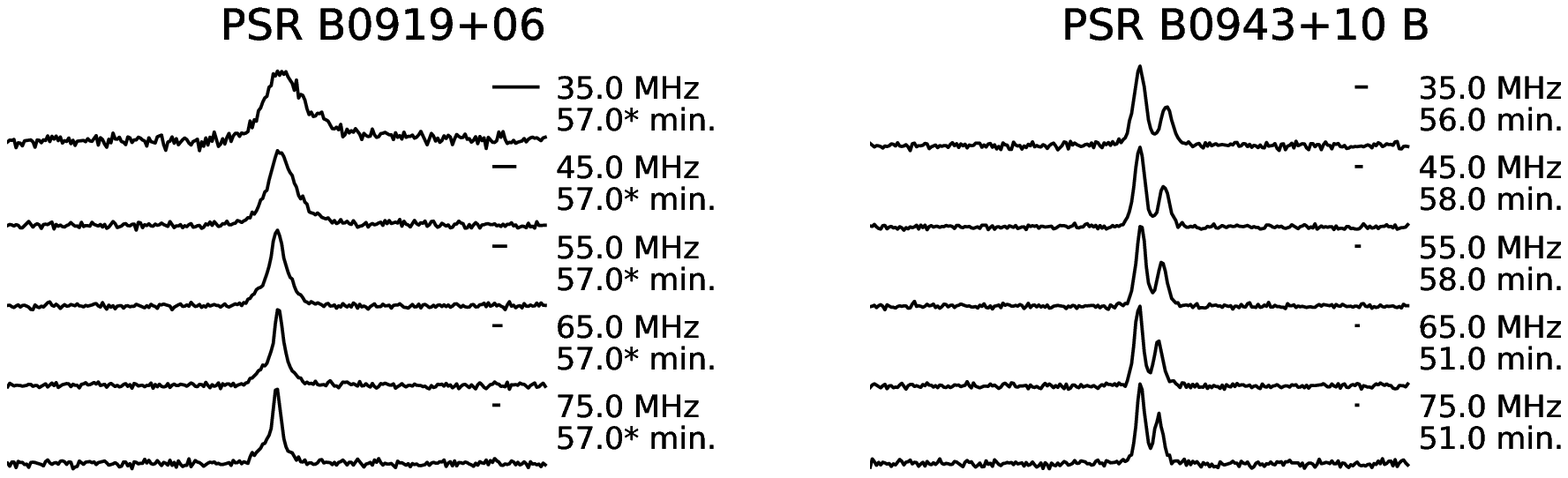}
\end{figure}
\begin{figure}
        \includegraphics[trim = 0mm 5mm 0mm 0mm, clip,width=0.5\textwidth]{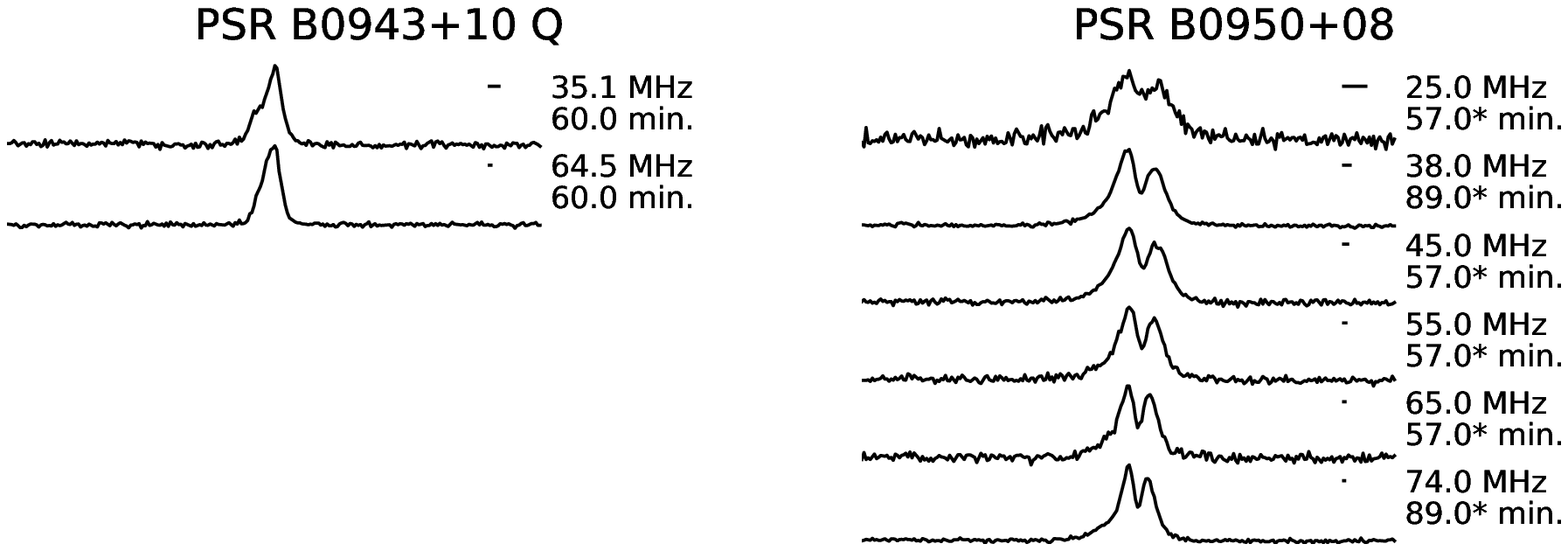}
\end{figure}
\begin{figure}
        \includegraphics[trim = 0mm 25mm 0mm 0mm, clip,width=0.5\textwidth]{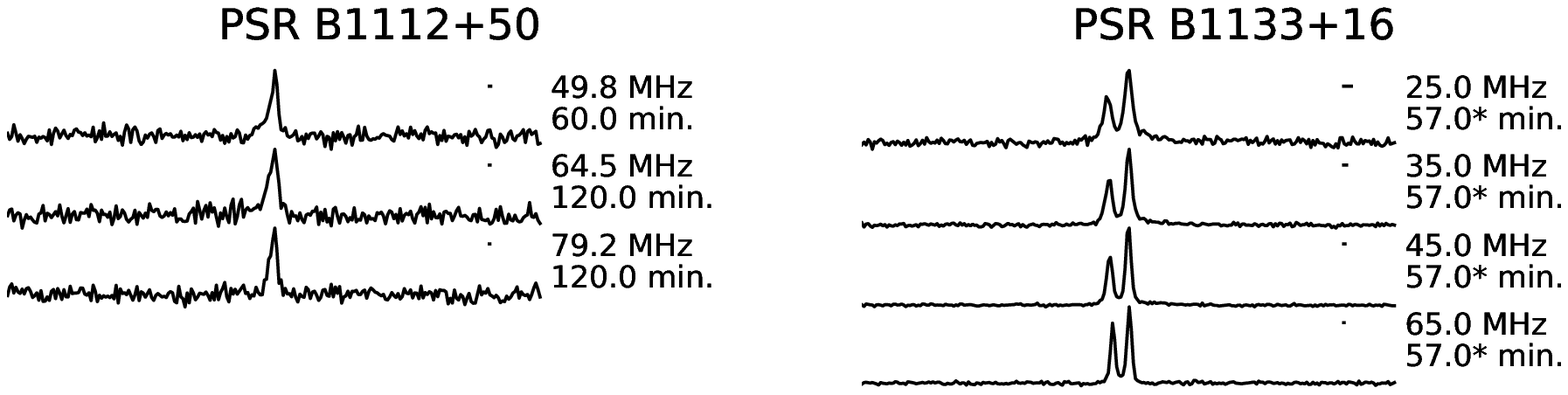}
\end{figure}
\begin{figure}
        \includegraphics[trim = 0mm 25mm 0mm 0mm, clip,width=0.5\textwidth]{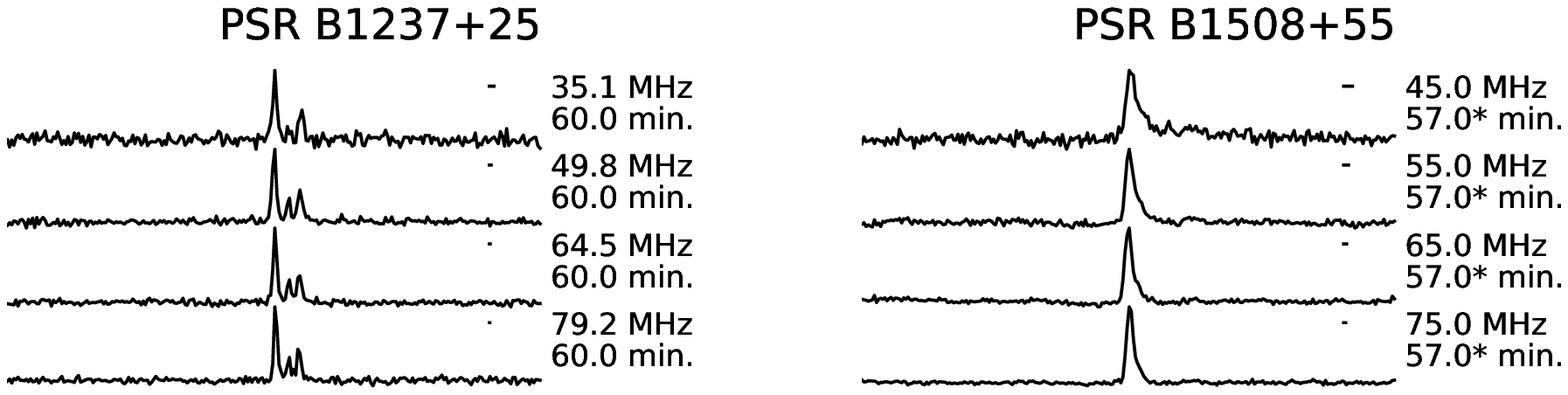}
\end{figure}
\begin{figure}
        \includegraphics[trim = 0mm 45mm 0mm 0mm, clip,width=0.5\textwidth]{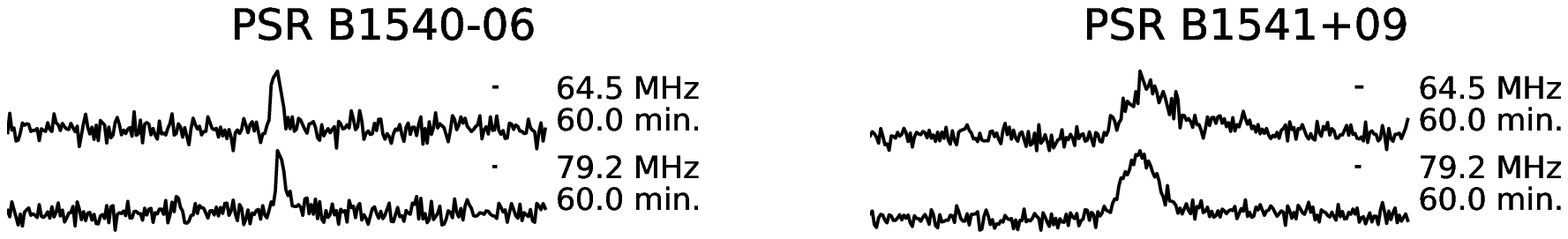}
\end{figure}
\begin{figure}
        \includegraphics[trim = 0mm 25mm 0mm 0mm, clip,width=0.5\textwidth]{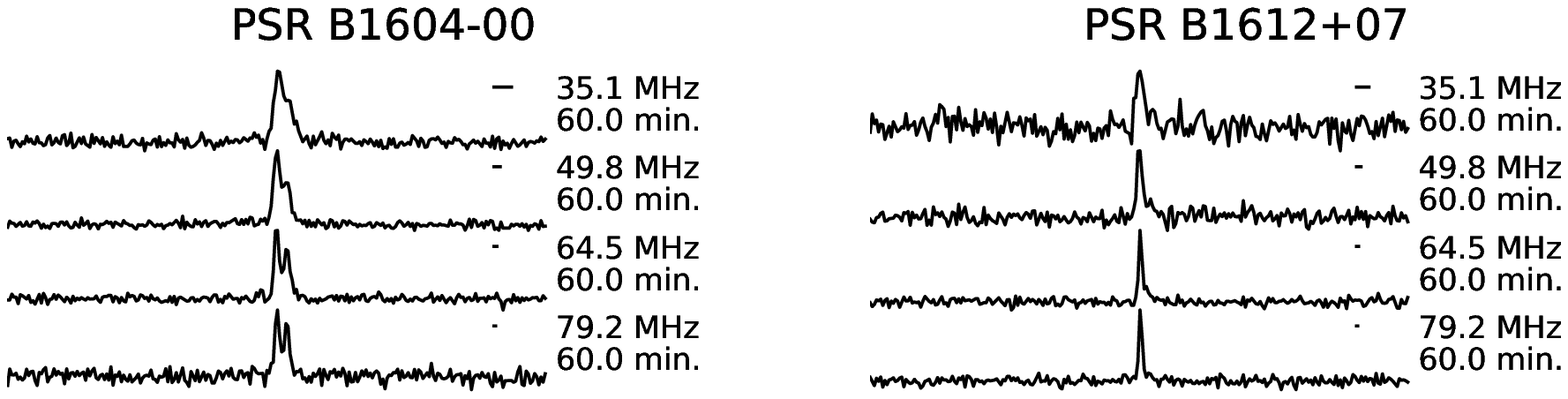}
\end{figure}
\begin{figure}
        \includegraphics[trim = 0mm 35mm 0mm 0mm, clip,width=0.5\textwidth]{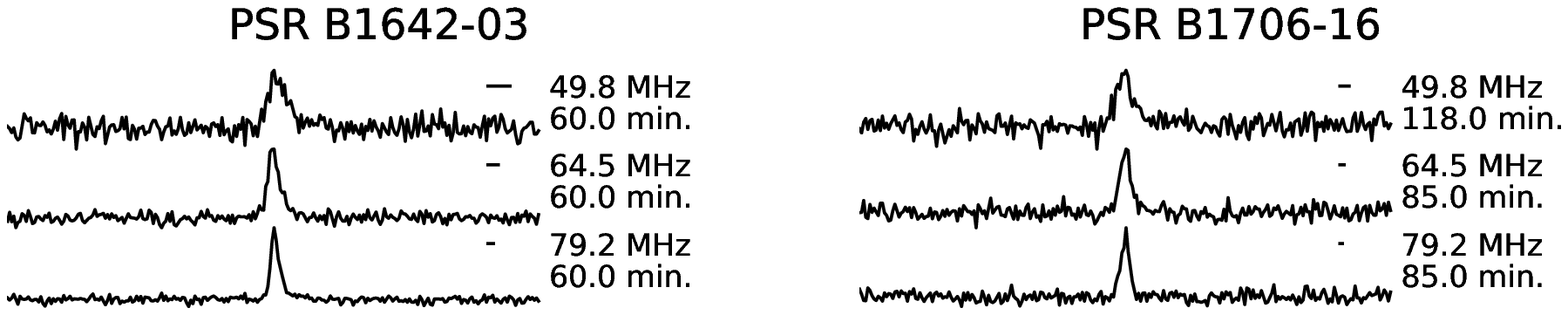}
\end{figure}
\begin{figure}
        \includegraphics[trim = 0mm 25mm 0mm 0mm, clip,width=0.5\textwidth]{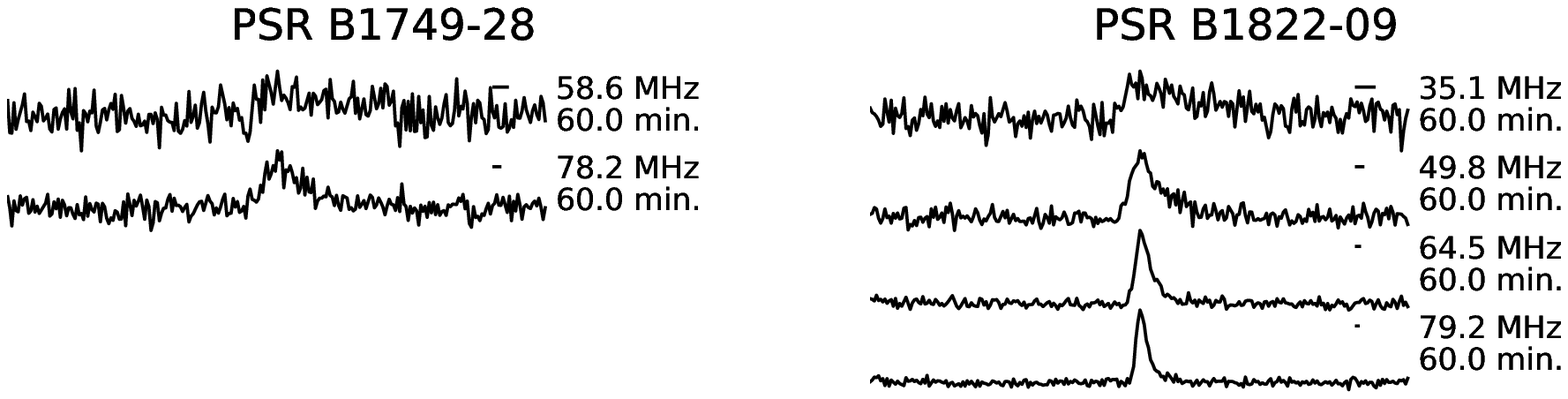}
\end{figure}
\begin{figure}
        \includegraphics[trim = 0mm 25mm 0mm 0mm, clip,width=0.5\textwidth]{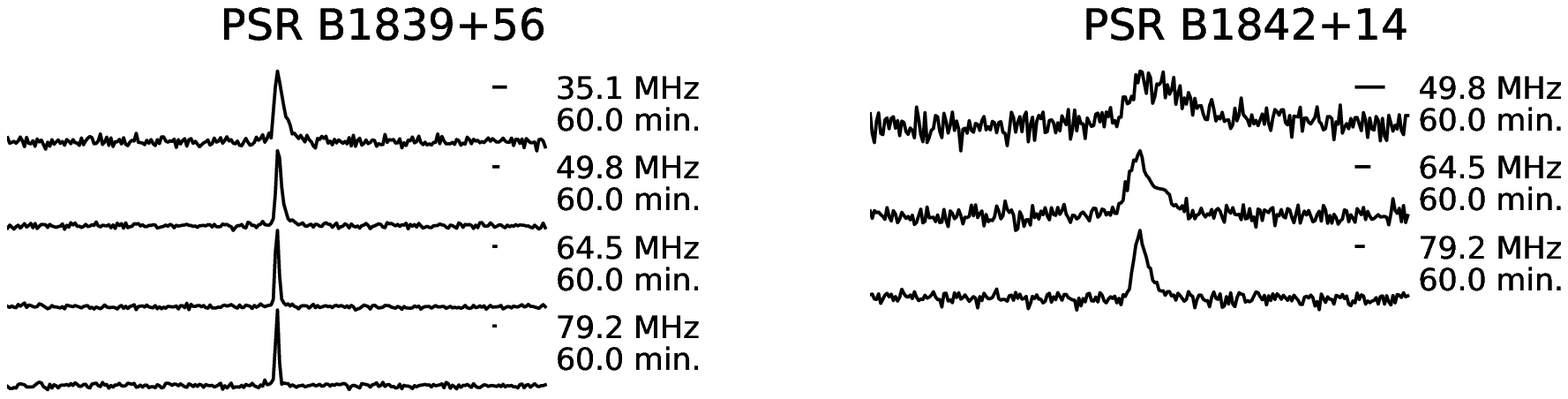}
\end{figure}
\begin{figure}
        \includegraphics[trim = 0mm 15mm 0mm 0mm, clip,width=0.5\textwidth]{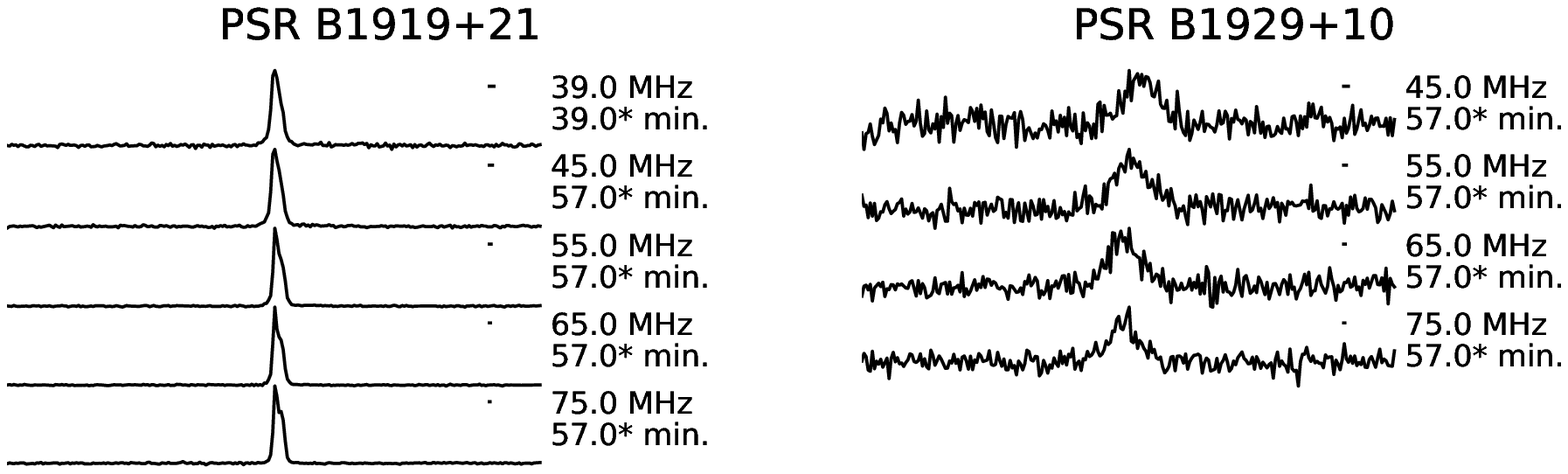}
\end{figure}
\begin{figure}
        \includegraphics[trim = 0mm 35mm 0mm 0mm, clip,width=0.5\textwidth]{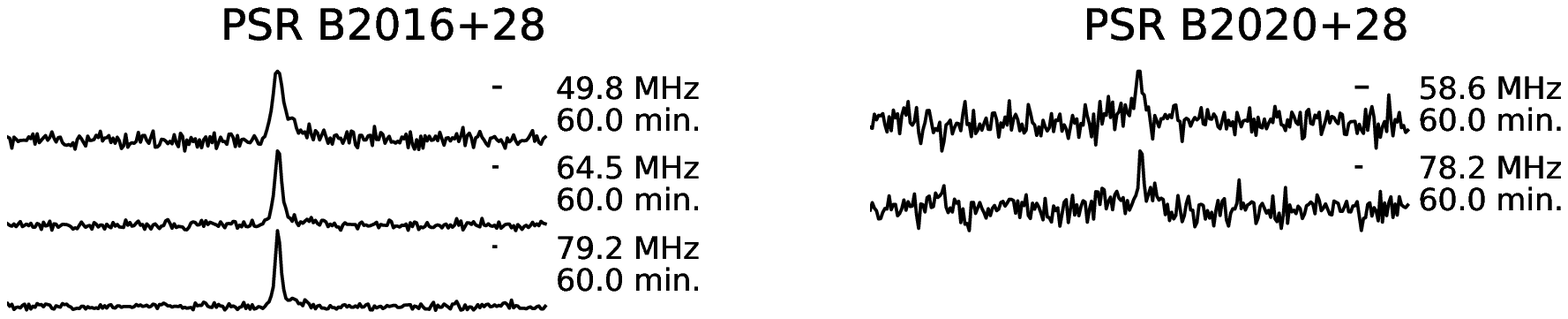}
\end{figure}
\begin{figure}
        \includegraphics[trim = 0mm 25mm 0mm 0mm, clip,width=0.5\textwidth]{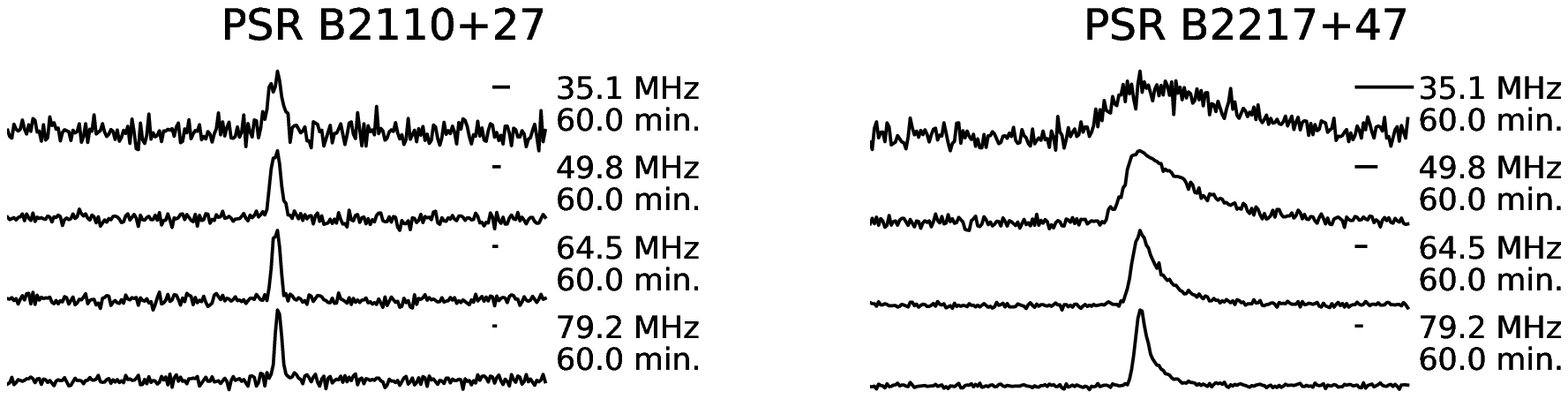}
\end{figure}
\begin{figure}
        \includegraphics[trim = 0mm 25mm 0mm 0mm, clip,width=0.5\textwidth]{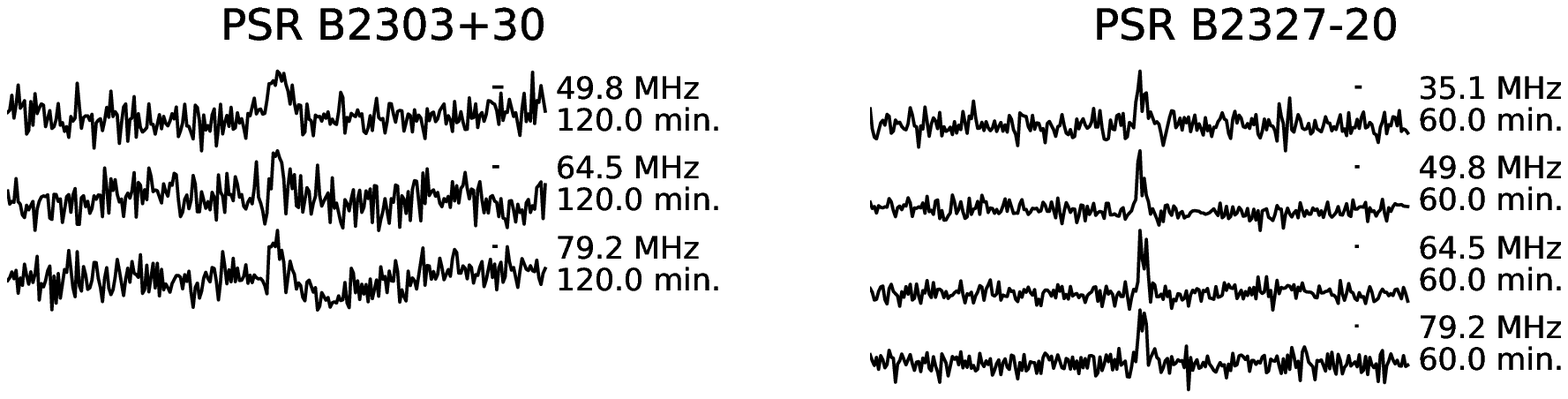}
\end{figure}
\begin{figure}
        \includegraphics[trim = 0mm 35mm 0mm 0mm, clip,width=0.5\textwidth]{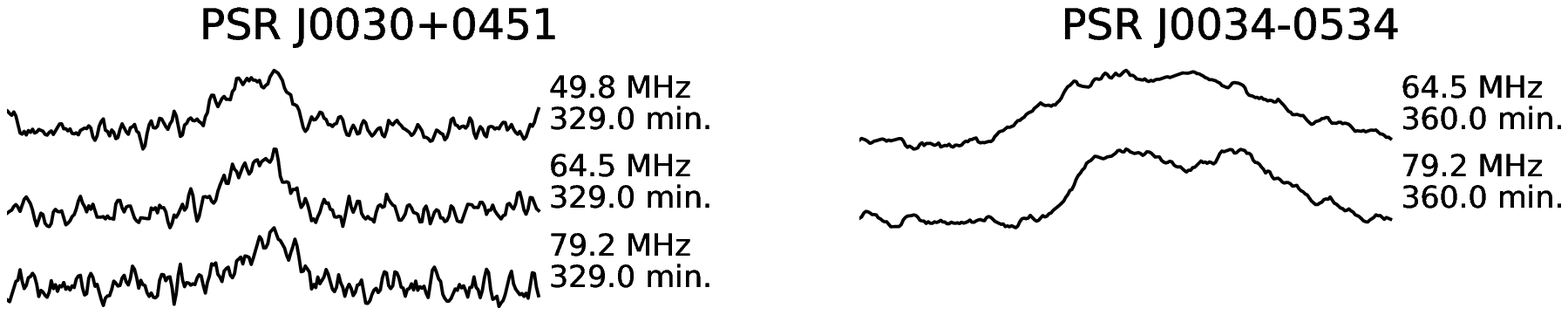}
\end{figure}
\begin{table*}[H]
\begin{center} {\footnotesize
\begin{tabular}{cccccccc}
\tablecaption{LWA1 Detected Pulsars}
\label{table:detections}
Pulsar & P & $\mathrm{DM_{ATNF}}$ & $\mathrm{DM_{Epoch}^{ATNF}}$ & $\mathrm{DM_{LWA1}}$ & $\mathrm{DM_{Epoch}^{LWA1}}$ & $\mathrm{DM_{UTR2}}$ & $\mathrm{DM_{Epoch}^{UTR2}}$\\
 & s & $\mathrm{pc\; cm^{-3}}$ & & $\mathrm{pc\; cm^{-3}}$ & & $\mathrm{pc\; cm^{-3}}$ & \\
\hline
J0030+0451 & 0.0049 & 4.333(1) & 50984 & 4.33252(4) & 56560 &   &  \\ 
B0031-07 & 0.9430 & 11.38(8) & 46635 & 10.922(6) & 56843 & 10.896(4) & 55480\\ 
J0034-0534 & 0.0019 & 13.76517(4) & 50690 & 13.76505(4) & 56631 &   &  \\ 
B0138+59 & 1.2229 & 34.797(11) & 49293 & 34.926(4) & 56563 &   &  \\ 
B0149-16 & 0.8327 & 11.922(4) & 48227 & 11.92577(4) & 56955 &   &  \\ 
B0301+19 & 1.3876 & 15.737(9) & 49289 & 15.68(3) & 57035 &   &  \\ 
B0320+39 & 3.0321 & 26.01(3) & 49290 & 26.173(2) & 56563 & 26.162(11) & 55480\\ 
B0329+54 & 0.7145 & 26.7641(1) & 46473 & 26.779(1) & 56364 & 25.661(11) & 55480\\ 
B0355+54 & 0.1564 & 57.1420(3) & 49616 & 57.1453(8) & 56674 &   &  \\ 
B0450+55 & 0.3407 & 14.495(7) & 49910 & 14.5943(9) & 56564 & 14.602(5) & 55480\\ 
B0525+21 & 3.7455 & 50.937(17) & 54200 & 50.93(1) & 56565 &   &  \\ 
B0628-28 & 1.2444 & 34.468(17) & 46603 & 34.425(1) & 56706 &   &  \\ 
B0655+64 & 0.1957 & 8.771(5) & 48806 & 8.777(1) & 56639 &   &  \\ 
B0809+74 & 1.2922 & 5.733(1) & 49162 & 5.771(2) & 56285 & 5.755(3) & 55480\\ 
B0818-13 & 1.2381 & 40.938(3) & 48904 & 40.981(2) & 56568 &   &  \\ 
B0823+26 & 0.5307 & 19.454(4) & 46450 & 19.4789(2) & 56665 & 19.484(6) & 55480\\ 
B0834+06 & 1.2738 & 12.889(6) & 48721 & 12.8640(4) & 56864 & 12.872(4) & 55480\\ 
B0906-17 & 0.4016 & 15.888(3) & 48737 & 15.879(2) & 56980 &   &  \\ 
B0919+06 & 0.4306 & 27.271(6) & 55140 & 27.2986(5) & 56285 & 27.316(6) & 55480\\ 
B0943+10 & 1.0977 & 15.4(5) & 48483 & 15.334(1) & 56365 & 15.339(4) & 55480\\ 
B0950+08 & 0.2531 & 2.958(3) & 46375 & 2.96927(8) & 56285 & 2.972(2) & 55480\\ 
B1112+50 & 1.6564 & 9.195(8) & 49334 & 9.1830(4) & 56687 & 9.185(4) & 55480\\ 
B1133+16 & 1.1879 & 4.8451(1) & 46407 & 4.8480(2) & 56743 & 4.846(2) & 55480\\ 
B1237+25 & 1.3824 & 9.242(6) & 46531 & 9.2575(3) & 56864 & 9.268(2) & 55480\\ 
B1508+55 & 0.7397 & 19.613(20) & 49904 & 19.6191(3) & 56284 & 19.622(9) & 55480\\ 
B1540-06 & 0.7091 & 18.403(4) & 49423 & 18.3774(9) & 56842 & 18.334(100) & 55480\\ 
B1541+09 & 0.7484 & 35.24(3) & 48716 & 35.012(5) & 56842 &   &  \\ 
B1604-00 & 0.4218 & 10.682(5) & 46973 & 10.6823(1) & 56843 & 10.688(2) & 55480\\ 
B1612+07 & 1.2068 & 21.39(3) & 49897 & 21.3949(3) & 56843 & 21.402(14) & 55480\\ 
B1642-03 & 0.3877 & 35.727(3) & 46515 & 35.7555(8) & 56842 &   &  \\ 
B1706-16 & 0.6531 & 24.873(5) & 46993 & 24.891(1) & 56843 &   &  \\ 
B1749-28 & 0.5626 & 50.372(8) & 46483 & 50.39(1) & 56568 &   &  \\ 
B1822-09 & 0.7690 & 19.38(4) & 54262 & 19.3833(9) & 56843 & 19.408(21) & 55480\\ 
B1839+56 & 1.6529 & 26.698(11) & 48717 & 26.774(1) & 56843 & 26.804(6) & 55480\\ 
B1842+14 & 0.3755 & 41.510(4) & 49362 & 41.498(1) & 56860 &   &  \\ 
B1919+21 & 1.3373 & 12.4370(1) & 48999 & 12.4386(3) & 56830 & 12.435(4) & 55480\\ 
B1929+10 & 0.2265 & 3.180(4) & 46523 & 3.1828(5) & 56294 & 3.180(2) & 55480\\ 
B2016+28 & 0.5580 & 14.172(4) & 46384 & 14.1977(6) & 57009 & 14.200(7) & 55480\\ 
B2020+28 & 0.3434 & 24.640(3) & 49692 & 24.632(1) & 56567 &   &  \\ 
B2110+27 & 1.2029 & 25.113(4) & 48741 & 25.1171(2) & 56842 & 25.114(18) & 55480\\ 
J2145-0750 & 0.0161 & 8.9977(14) & 53040 & 9.00470(9) & 56403 &   &  \\ 
B2217+47 & 0.5385 & 43.519(12) & 46599 & 43.4975(5) & 56842 &   &  \\ 
B2303+30 & 1.5759 & 49.544(16) & 48714 & 49.639(6) & 57048 &   &  \\ 
B2327-20 & 1.6436 & 8.458(13) & 49878 & 8.456(2) & 56979 &   &  \\ 
\hline
\end{tabular}
\caption[DMs for LWA1-detected pulsars.]{Dispersion Measures for 44 pulsars detected by LWA1. We also list the DM values reported in the ATNF catalog as well as values reported in ~\citep{2013MNRAS.431.3624Z}}}
\end{center}
\end{table*}

\begin{figure}
        \includegraphics[trim = 0mm 15mm 0mm 0mm, clip,width=0.5\textwidth]{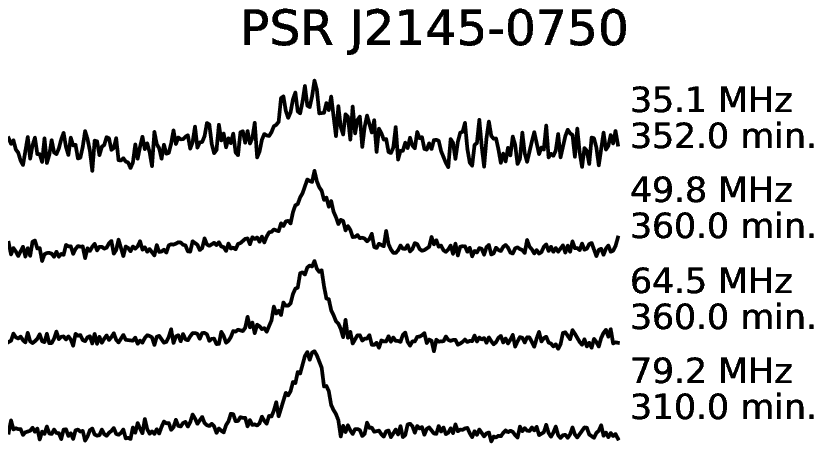}
\caption{Integrated pulse profiles for {\bfref 44} pulsars detected with LWA1 at a variety of observing
frequencies. In many cases, profiles at different frequencies were obtained at different times
and reduced in different ways, so they have been aligned manually. Profiles for PSRs J0030+0451,
J0034-0534, and J2145-0750 were coherently de-dispersed, while all others were incoherently
de-dispersed. Incoherently de-dispersed profiles have a line showing the dispersive smearing
time within the center frequency channel relative to the pulse period. Each profile contains
256 pulse phase bins. The center frequency and total amount of integration time is shown to the
right of each profile. Integration times with a * are from observations obtained prior to 28 February
2013, when the telescope had a less optimal cable delay calibration. {\bfref All profiles are plotted
over the full phase of the pulsar.}}
\label{fig:prof1}
\end{figure}

{ \footnotesize
\begin{longtable*}[H]{ccccccccc}
Pulsar & $\nu$ & T & $\mathrm{w_{50}}$ & $\mathrm{w_{10}}$ & SNR & SEFD & $\mathrm{S_{\nu}}$ & $\alpha$ \\
 & MHz & min. & phase & phase & & kJy & mJy & \\
\hline\hline
\endhead
B0031-07 & 35.1 & 60 & 0.074 & 0.242 & 196.29 & 14.88 & 3490(1740) & 0.5(2)\\
B0031-07 & 49.8 & 60 & 0.137 & 0.246 & 318.75 & 11.91 & 6370(3190) & \\
B0031-07 & 64.5 & 60 & 0.160 & 0.223 & 331.25 & 11.91 & 7260(3630) & \\
B0031-07 & 79.2 & 60 & 0.121 & 0.188 & 271.99 & 11.91 & 5070(2540) & \\
\hline
B0138+59 & 58.6 & 60 & 0.027 & 0.051 & 73.65 & 8.89 & 460(230) & \\
B0138+59 & 78.2 & 60 & 0.020 & 0.043 & 80.22 & 8.89 & 420(210) & \\
\hline
B0149-16 & 35.1 & 60 & 0.027 & 0.047 & 59.76 & 18.85 & 800(400) & -1.6(1) \\
B0149-16 & 49.8 & 60 & 0.012 & 0.043 & 82.86 & 15.09 & 570(290) & \\
B0149-16 & 64.5 & 60 & 0.012 & 0.027 & 69.72 & 15.09 & 480(240) & \\
B0149-16 & 79.2 & 60 & 0.004 & 0.031 & 49.54 & 15.09 & 200(100) & \\
\hline
B0301+19 & 64.5 & 120 & 0.023 & 0.039 & 35.85 & 7.47 & 120(60) & \\
B0301+19 & 79.2 & 120 & 0.020 & 0.043 & 42.64 & 7.47 & 130(70) & \\
\hline
B0320+39 & 39.0 & 60 & 0.020 & 0.062 & 43.51 & 8.37 & 220(110) & \\
B0320+39 & 58.6 & 60 & 0.020 & 0.035 & 58.28 & 6.70 & 230(120) & \\
\hline
B0329+54 & 35.1 & 29 & 0.070 & 0.191 & 186.06 & 10.94 & 3370(1680) & 0.00(9) \\
B0329+54 & 45.0 & 60 & 0.066 & 0.195 & 422.49 & 8.15 & 3870(1940) & \\
B0329+54 & 49.8 & 29 & 0.035 & 0.129 & 414.30 & 8.75 & 4160(2080) & \\
B0329+54 & 55.0 & 60 & 0.031 & 0.102 & 541.62 & 8.15 & 3340(1670) & \\
B0329+54 & 65.0 & 60 & 0.023 & 0.078 & 574.76 & 8.15 & 3060(1530) & \\
B0329+54 & 74.0 & 13 & 0.016 & 0.051 & 354.11 & 10.16 & 4070(2030) & \\
\hline
B0355+54 & 79.2 & 60 & 0.059 & 0.113 & 52.91 & 8.10 & 450(230) & \\
\hline
B0450+55 & 39.0 & 60 & 0.059 & 0.113 & 64.03 & 10.34 & 700(350) & -1.3(2) \\
B0450+55 & 58.6 & 59 & 0.023 & 0.117 & 126.84 & 8.28 & 690(350) & \\
B0450+55 & 78.2 & 59 & 0.020 & 0.043 & 56.94 & 8.28 & 280(140) & \\
\hline
B0525+21 & 39.0 & 60 & 0.055 & 0.105 & 106.63 & 9.04 & 980(490) & -0.8(1) \\
B0525+21 & 58.6 & 60 & 0.027 & 0.059 & 109.17 & 7.24 & 560(280) & \\
B0525+21 & 78.2 & 60 & 0.023 & 0.039 & 123.62 & 7.24 & 580(290) & \\
\hline
B0628-28 & 35.1 & 60 & 0.082 & 0.152 & 125.32 & 28.71 & 4540(2270) & 0.40(3) \\
B0628-28 & 49.8 & 60 & 0.074 & 0.137 & 193.16 & 22.97 & 5300(2650) & \\
B0628-28 & 64.5 & 60 & 0.059 & 0.113 & 223.86 & 22.97 & 5410(2710) & \\
B0628-28 & 79.2 & 60 & 0.059 & 0.113 & 271.47 & 22.97 & 6560(3280) & \\
\hline
B0655+64 & 49.8 & 49 & 0.043 & 0.082 & 67.75 & 9.52 & 640(320) & -0.7(1) \\
B0655+64 & 64.5 & 49 & 0.039 & 0.055 & 70.96 & 9.52 & 640(320) & \\
B0655+64 & 79.2 & 49 & 0.035 & 0.059 & 53.58 & 9.52 & 460(230) & \\
\hline
B0818-13 & 58.6 & 60 & 0.020 & 0.043 & 44.50 & 14.04 & 370(190) & \\
B0818-13 & 78.2 & 60 & 0.020 & 0.035 & 71.46 & 14.04 & 600(300) & \\
\hline
B0823+26 & 64.5 & 119 & 0.020 & 0.043 & 111.52 & 6.84 & 450(230) & \\
B0823+26 & 79.2 & 119 & 0.020 & 0.035 & 17.16 & 6.84 & 70(30) & \\
\hline
B0834+06 & 35.1 & 60 & 0.027 & 0.051 & 345.45 & 11.44 & 2790(1400) & 0.5(1) \\
B0834+06 & 49.8 & 60 & 0.027 & 0.047 & 723.39 & 9.16 & 4680(2340) & \\
B0834+06 & 64.5 & 60 & 0.023 & 0.039 & 800.31 & 9.16 & 4790(2390) & \\
B0834+06 & 79.2 & 60 & 0.023 & 0.031 & 667.24 & 9.16 & 3990(2000) & \\
\hline
B0906-17 & 49.8 & 60 & 0.051 & 0.090 & 47.71 & 15.61 & 730(360) & -1.65(1) \\
B0906-17 & 64.5 & 60 & 0.027 & 0.051 & 42.61 & 15.61 & 470(240) & \\
B0906-17 & 79.2 & 60 & 0.027 & 0.051 & 30.45 & 15.61 & 340(170) & \\
\hline
B1112+50 & 35.1 & 60 & 0.020 & 0.027 & 27.34 & 7.66 & 120(60) & 0.7(2) \\
B1112+50 & 49.8 & 60 & 0.016 & 0.051 & 76.58 & 7.66 & 310(160) & \\
B1112+50 & 64.5 & 60 & 0.020 & 0.035 & 55.09 & 7.66 & 250(130) & \\
B1112+50 & 79.2 & 60 & 0.020 & 0.035 & 44.34 & 7.66 & 200(100) & \\
\hline
B1133+16 & 35.1 & 60 & 0.023 & 0.066 & 600.98 & 9.84 & 3850(1930) & 0.30(5) \\
B1133+16 & 45.0 & 100 & 0.020 & 0.059 & 1024.48 & 8.55 & 4040(2020) & \\
B1133+16 & 49.8 & 60 & 0.020 & 0.059 & 1008.95 & 7.88 & 4710(2360) & \\
B1133+16 & 64.5 & 60 & 0.020 & 0.051 & 1096.32 & 7.88 & 5120(2560) & \\
B1133+16 & 79.2 & 60 & 0.020 & 0.047 & 1012.97 & 7.88 & 4730(2370) & \\
\hline
B1237+25 & 35.1 & 60 & 0.012 & 0.043 & 69.96 & 8.73 & 280(140) & 0.0(2) \\
B1237+25 & 49.8 & 60 & 0.012 & 0.047 & 158.29 & 6.98 & 510(250) & \\
B1237+25 & 64.5 & 60 & 0.012 & 0.039 & 160.37 & 6.98 & 510(260) & \\
B1237+25 & 79.2 & 60 & 0.004 & 0.039 & 141.41 & 6.98 & 260(130) & \\
\hline
B1508+55 & 58.6 & 14 & 0.016 & 0.047 & 236.95 & 8.24 & 2130(1060) & \\
B1508+55 & 78.2 & 14 & 0.016 & 0.039 & 285.24 & 8.24 & 2560(1280) & \\
\hline
B1540-06 & 49.8 & 60 & 0.020 & 0.035 & 38.91 & 11.67 & 270(140) & -1.7(4) \\
B1540-06 & 64.5 & 60 & 0.020 & 0.035 & 46.68 & 11.67 & 320(160) & \\
B1540-06 & 79.2 & 60 & 0.020 & 0.035 & 17.18 & 11.67 & 120(60) & \\
\hline
B1541+09 & 64.5 & 60 & 0.098 & 0.184 & 167.85 & 8.67 & 2020(1010) & \\
B1541+09 & 79.2 & 60 & 0.074 & 0.129 & 66.21 & 8.67 & 690(340) & \\
\hline
B1604-00 & 35.1 & 60 & 0.035 & 0.055 & 153.93 & 12.66 & 1570(780) & -1.5(1) \\
B1604-00 & 49.8 & 60 & 0.023 & 0.047 & 181.30 & 10.13 & 1200(600) & \\
B1604-00 & 64.5 & 60 & 0.023 & 0.043 & 149.07 & 10.13 & 990(490) & \\
B1604-00 & 79.2 & 60 & 0.020 & 0.039 & 75.03 & 10.13 & 450(230) & \\
\hline
B1612+07 & 35.1 & 60 & 0.020 & 0.027 & 30.53 & 11.16 & 200(100) & -0.72(6) \\
B1612+07 & 49.8 & 60 & 0.012 & 0.020 & 43.17 & 8.93 & 180(90) & \\
B1612+07 & 64.5 & 60 & 0.004 & 0.023 & 64.70 & 8.93 & 150(80) & \\
B1612+07 & 79.2 & 60 & 0.004 & 0.012 & 44.82 & 8.93 & 110(50) & \\
\hline
B1642-03 & 49.8 & 60 & 0.043 & 0.074 & 78.63 & 10.93 & 770(380) & 0.2(1) \\
B1642-03 & 64.5 & 60 & 0.027 & 0.051 & 120.04 & 10.93 & 930(460) & \\
B1642-03 & 79.2 & 60 & 0.020 & 0.035 & 126.84 & 10.93 & 830(410) & \\
\hline
B1706-16 & 49.8 & 60 & 0.035 & 0.059 & 73.83 & 15.15 & 900(450) & -0.65(5) \\
B1706-16 & 64.5 & 60 & 0.020 & 0.043 & 79.31 & 15.15 & 710(360) & \\
B1706-16 & 79.2 & 60 & 0.020 & 0.035 & 74.02 & 15.15 & 670(330) & \\
\hline
B1749-28 & 78.2 & 60 & 0.082 & 0.152 & 117.02 & 22.57 & 3330(1660) & \\
\hline
B1822-09 & 35.1 & 60 & 0.160 & 0.293 & 106.51 & 15.70 & 3080(1540) & -0.77(2) \\
B1822-09 & 49.8 & 60 & 0.047 & 0.141 & 189.29 & 12.56 & 2220(1110) & \\
B1822-09 & 64.5 & 60 & 0.027 & 0.070 & 209.32 & 12.56 & 1860(930) & \\
B1822-09 & 79.2 & 60 & 0.023 & 0.059 & 200.24 & 12.56 & 1640(820) & \\
\hline
B1839+56 & 35.1 & 60 & 0.020 & 0.047 & 136.16 & 10.50 & 850(430) & -1.7(2) \\
B1839+56 & 49.8 & 60 & 0.012 & 0.027 & 176.24 & 8.40 & 680(340) & \\
B1839+56 & 64.5 & 60 & 0.012 & 0.012 & 135.70 & 8.40 & 520(260) & \\
B1839+56 & 79.2 & 60 & 0.004 & 0.012 & 94.31 & 8.40 & 210(100) & \\
\hline
B1842+14 & 49.8 & 60 & 0.129 & 0.230 & 159.79 & 7.97 & 2070(1030) & -2.8(3) \\
B1842+14 & 64.5 & 60 & 0.047 & 0.117 & 178.22 & 7.97 & 1330(660) & \\
B1842+14 & 79.2 & 60 & 0.035 & 0.059 & 127.88 & 7.97 & 820(410) & \\
\hline
B1919+21 & 35.1 & 358 & 0.020 & 0.055 & 738.05 & 9.05 & 1620(810) & 0.17(2) \\
B1919+21 & 49.8 & 358 & 0.020 & 0.043 & 922.67 & 7.24 & 1630(810) & \\
B1919+21 & 64.5 & 358 & 0.020 & 0.035 & 979.29 & 7.24 & 1730(860) & \\
B1919+21 & 79.2 & 358 & 0.020 & 0.031 & 1055.24 & 7.24 & 1860(930) & \\
\hline
B2016+28 & 49.8 & 60 & 0.023 & 0.078 & 135.41 & 6.66 & 590(290) & -0.18(9) \\
B2016+28 & 64.5 & 60 & 0.020 & 0.035 & 127.30 & 6.66 & 500(250) & \\
B2016+28 & 79.2 & 60 & 0.012 & 0.039 & 178.14 & 6.66 & 550(270) & \\
\hline
B2020+28 & 58.6 & 60 & 0.020 & 0.035 & 28.37 & 6.65 & 110(60) & \\
B2020+28 & 78.2 & 60 & 0.012 & 0.020 & 19.26 & 6.65 & 60(30) & \\
\hline
B2110+27 & 35.1 & 60 & 0.035 & 0.059 & 60.18 & 8.44 & 410(200) & -0.5(1) \\
B2110+27 & 49.8 & 60 & 0.020 & 0.043 & 111.89 & 6.75 & 450(220) & \\
B2110+27 & 64.5 & 60 & 0.020 & 0.027 & 99.09 & 6.75 & 400(200) & \\
B2110+27 & 79.2 & 60 & 0.012 & 0.027 & 86.06 & 6.75 & 270(130) & \\
\hline
B2217+47 & 35.1 & 60 & 0.250 & 0.492 & 470.76 & 9.79 & 11220(5610) & -1.8(2) \\
B2217+47 & 49.8 & 60 & 0.121 & 0.316 & 823.71 & 7.83 & 10100(5050) & \\
B2217+47 & 64.5 & 60 & 0.051 & 0.156 & 732.30 & 7.83 & 5600(2800) & \\
B2217+47 & 79.2 & 60 & 0.023 & 0.082 & 492.77 & 7.83 & 2520(1260) & \\
\hline
B2303+30 & 49.8 & 120 & 0.043 & 0.074 & 42.24 & 6.65 & 180(90) & -3.3(6) \\
B2303+30 & 64.5 & 120 & 0.020 & 0.035 & 19.24 & 6.65 & 50(30) & \\
B2303+30 & 79.2 & 120 & 0.020 & 0.035 & 18.44 & 6.65 & 50(30) & \\
\hline
B2327-20 & 35.1 & 60 & 0.020 & 0.035 & 25.98 & 20.93 & 320(160) & -0.5(1) \\
B2327-20 & 49.8 & 60 & 0.004 & 0.004 & 69.63 & 16.75 & 310(150) & \\
B2327-20 & 64.5 & 60 & 0.004 & 0.016 & 46.02 & 16.75 & 200(100) & \\
B2327-20 & 79.2 & 60 & 0.016 & 0.020 & 27.87 & 16.75 & 250(120) & \\
\hline
\caption[Fluxes for LWA1-detected pulsars.]{Mean flux density ($\mathrm{S_\nu}$) measurements for {\bfref 36} pulsars detected {\bfref using} LWA1 {\bfref at multiple observing frequencies ($\mathrm{\nu}$ and measured spectral indices for pulsars with more than 2 measurements of $\mathrm{S_\nu}$}. {\bfref We also give the integration time (T) of the observations used in these measurements as well as pulse width at half-maximum ($\mathrm{w_{50}}$) and 10\% of maximum ($\mathrm{w_{10}}$).}}
\label{table:flux}
\end{longtable*} }

\begin{figure*}[h]
        \includegraphics[width=0.9\textwidth]{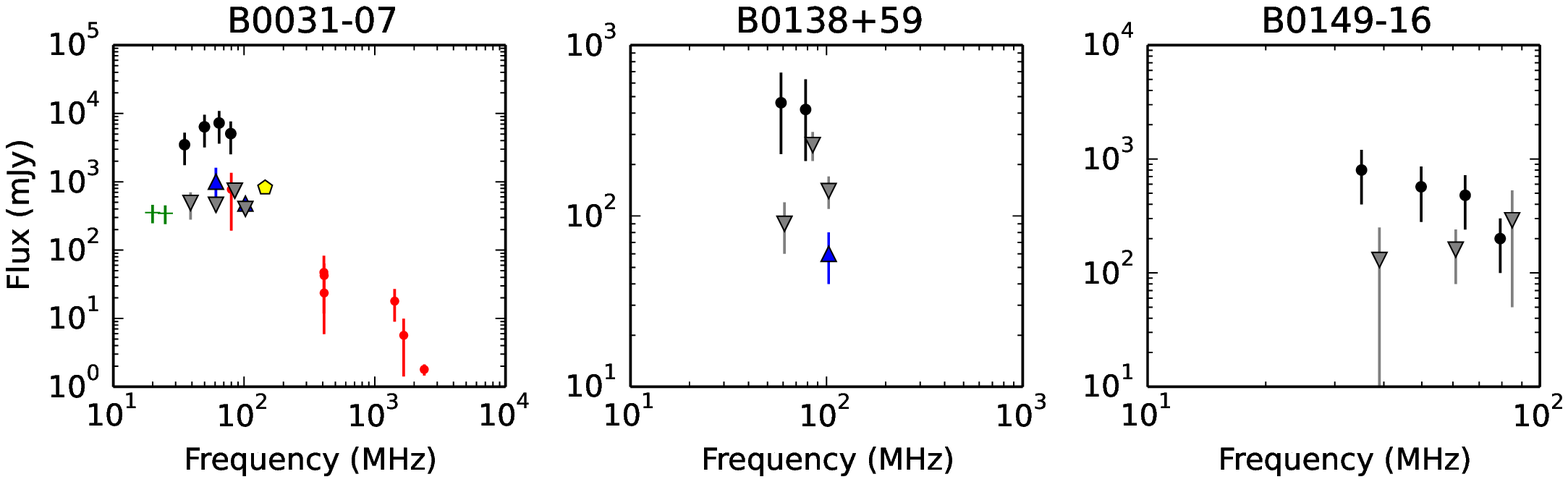}
        \includegraphics[width=0.9\textwidth]{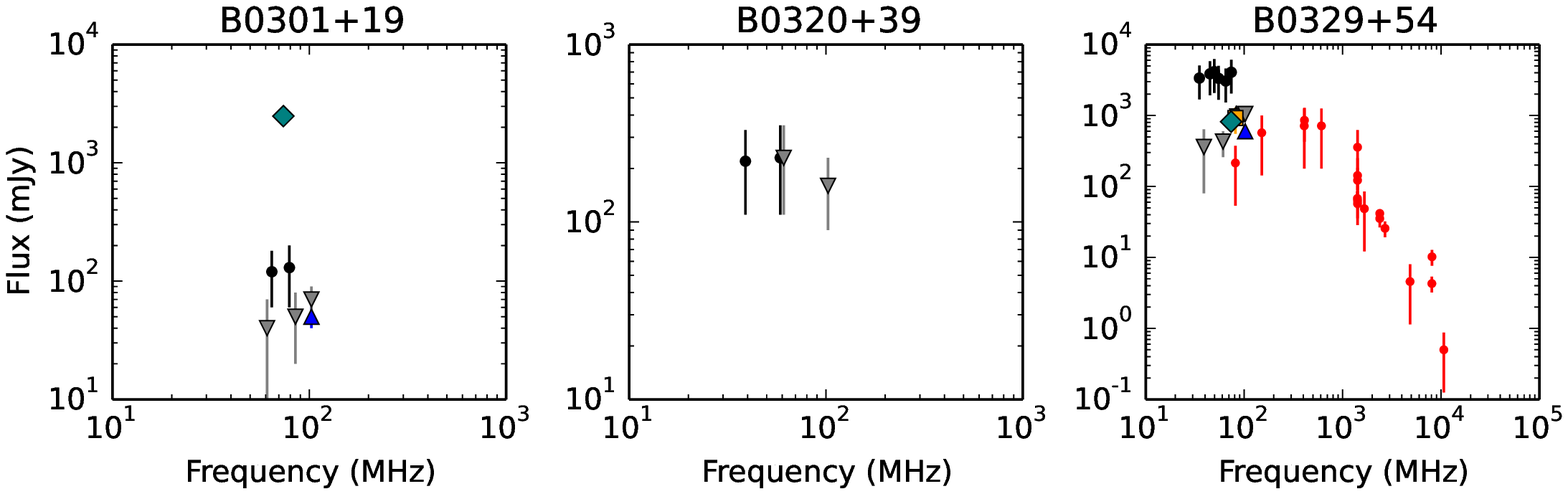}
        \includegraphics[width=0.9\textwidth]{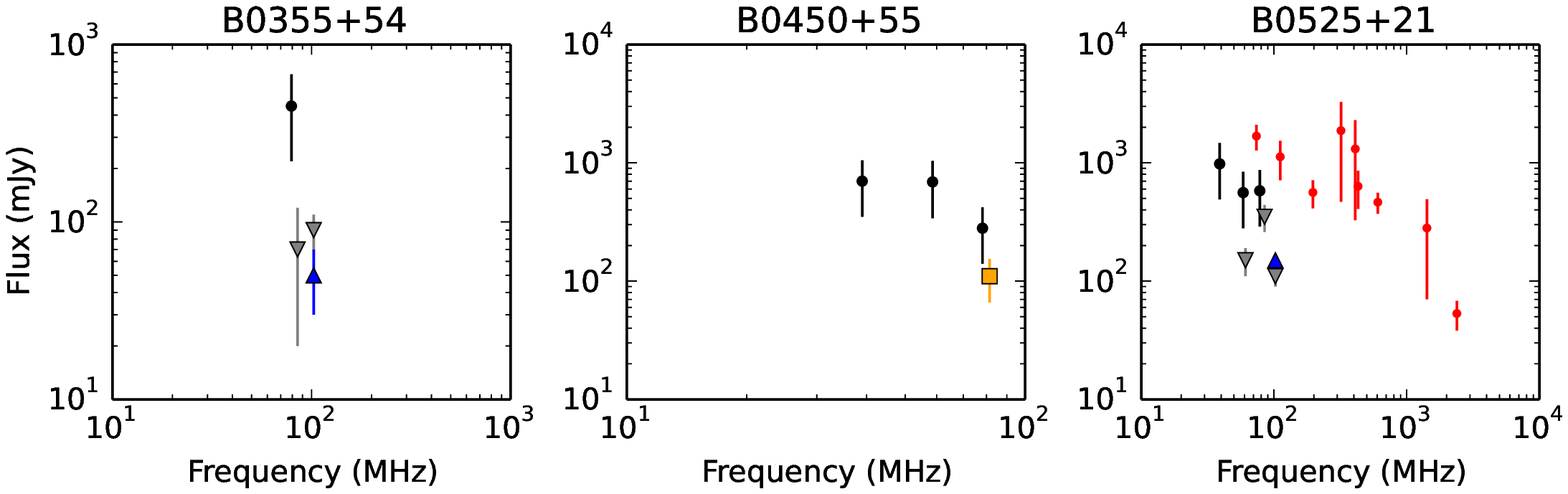}
        \includegraphics[width=0.9\textwidth]{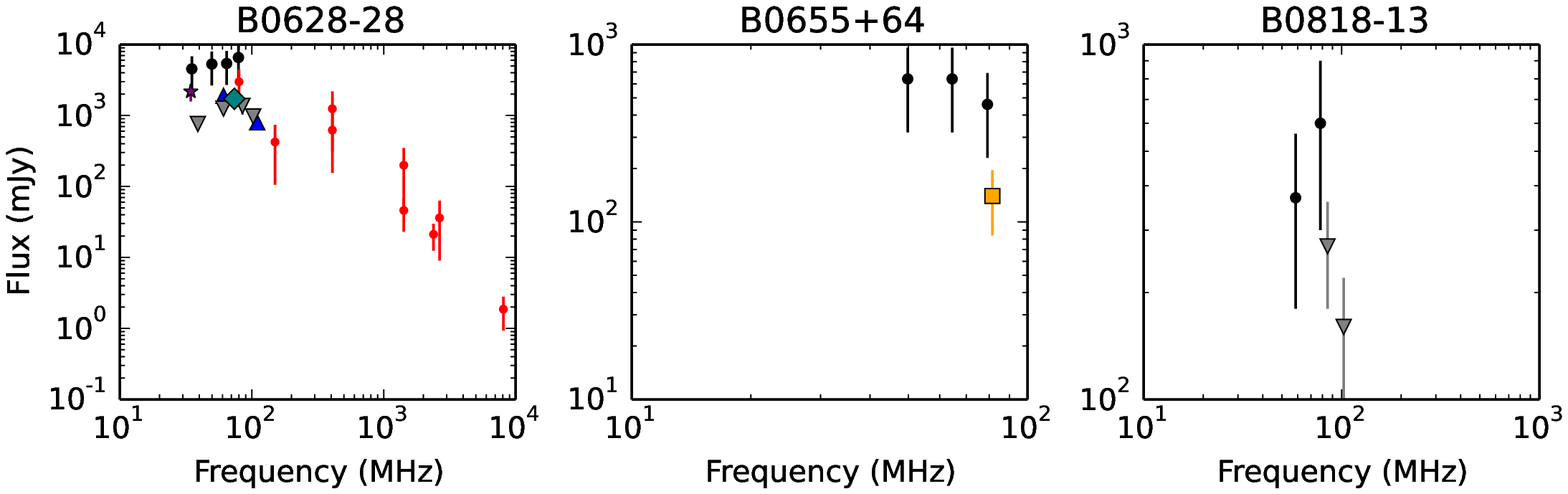}
\end{figure*}
\begin{figure*}[h]
        \includegraphics[width=0.9\textwidth]{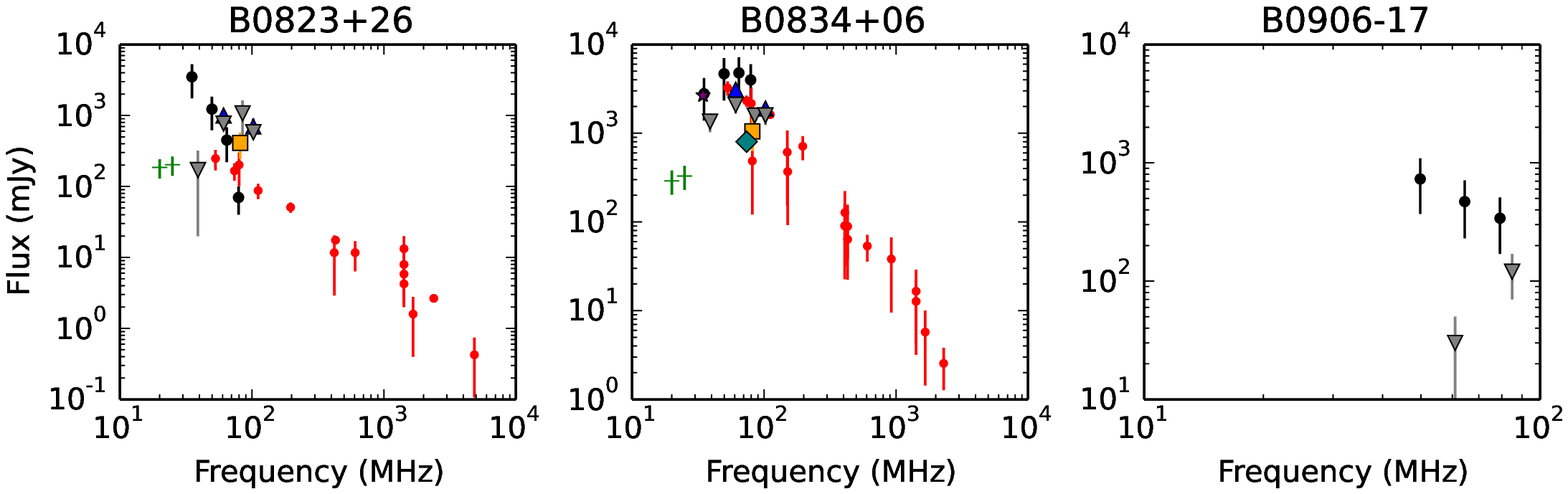}
        \includegraphics[width=0.9\textwidth]{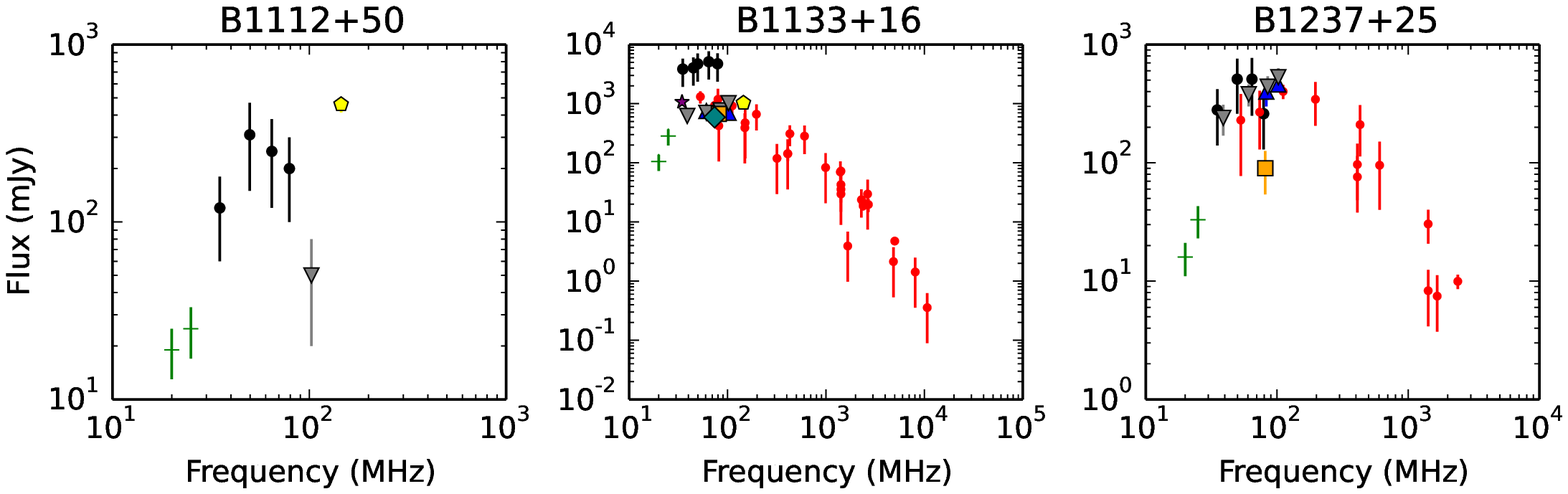}
        \includegraphics[width=0.9\textwidth]{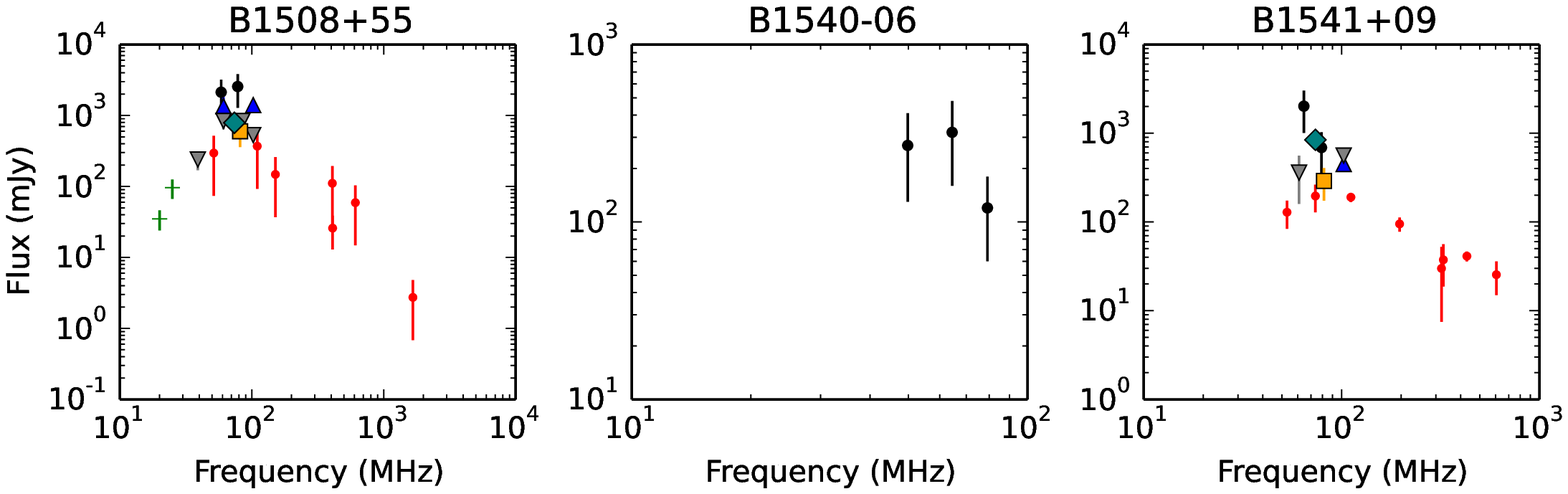}
        \includegraphics[width=0.9\textwidth]{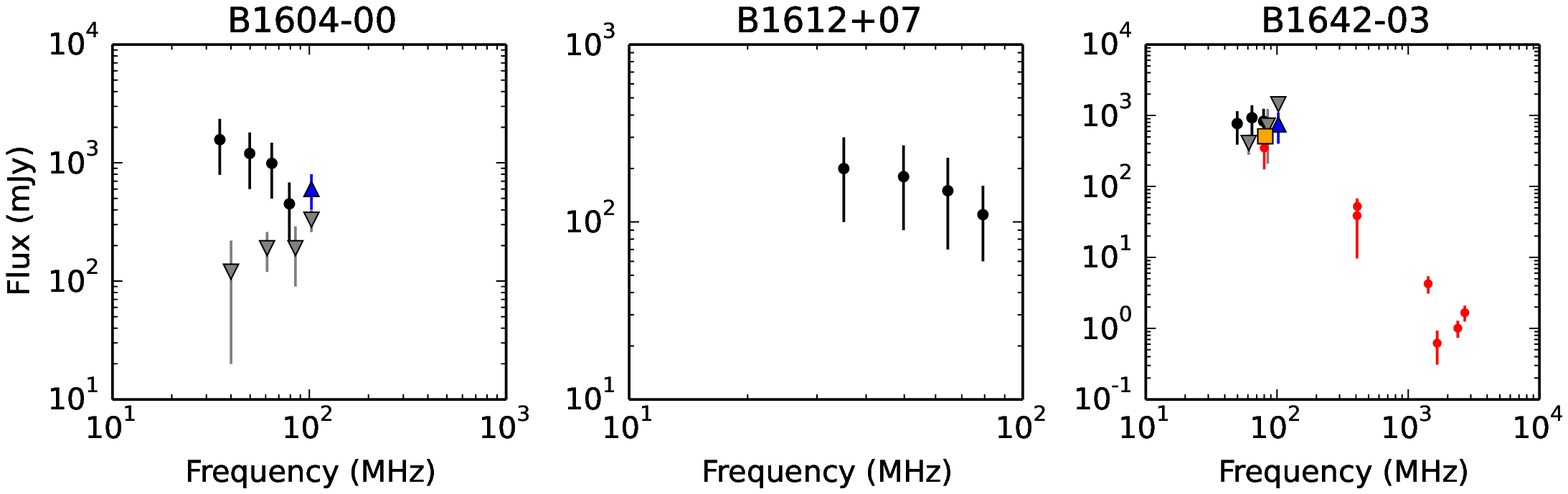}
\end{figure*}
\begin{figure*}[h]
        \includegraphics[width=0.9\textwidth]{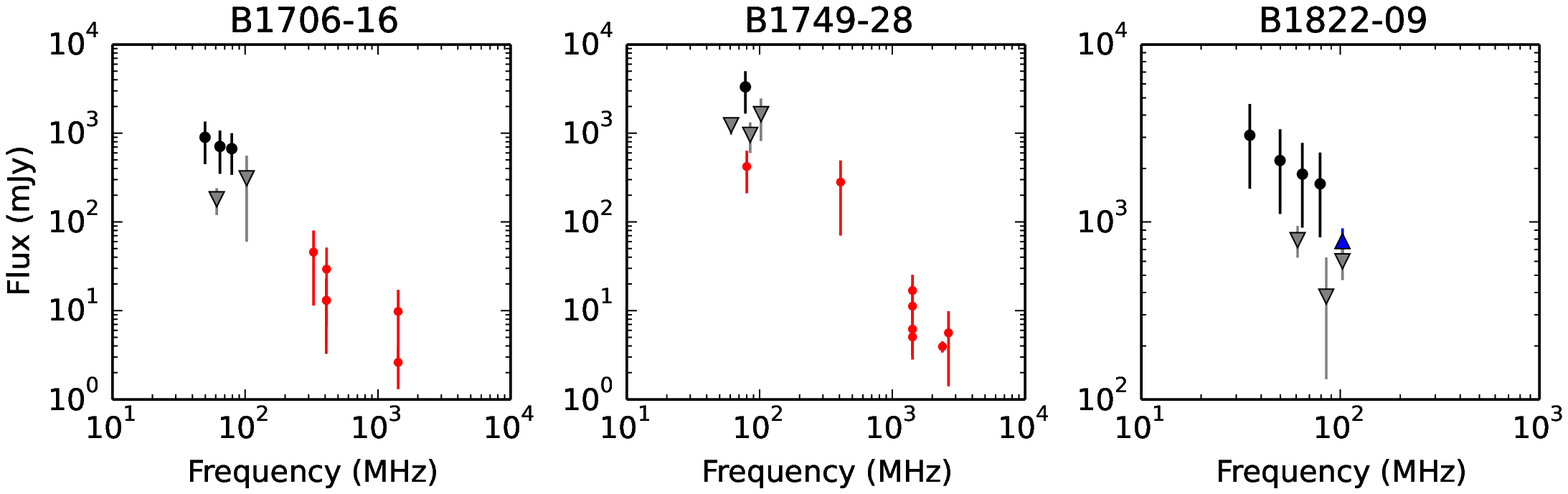}
        \includegraphics[width=0.9\textwidth]{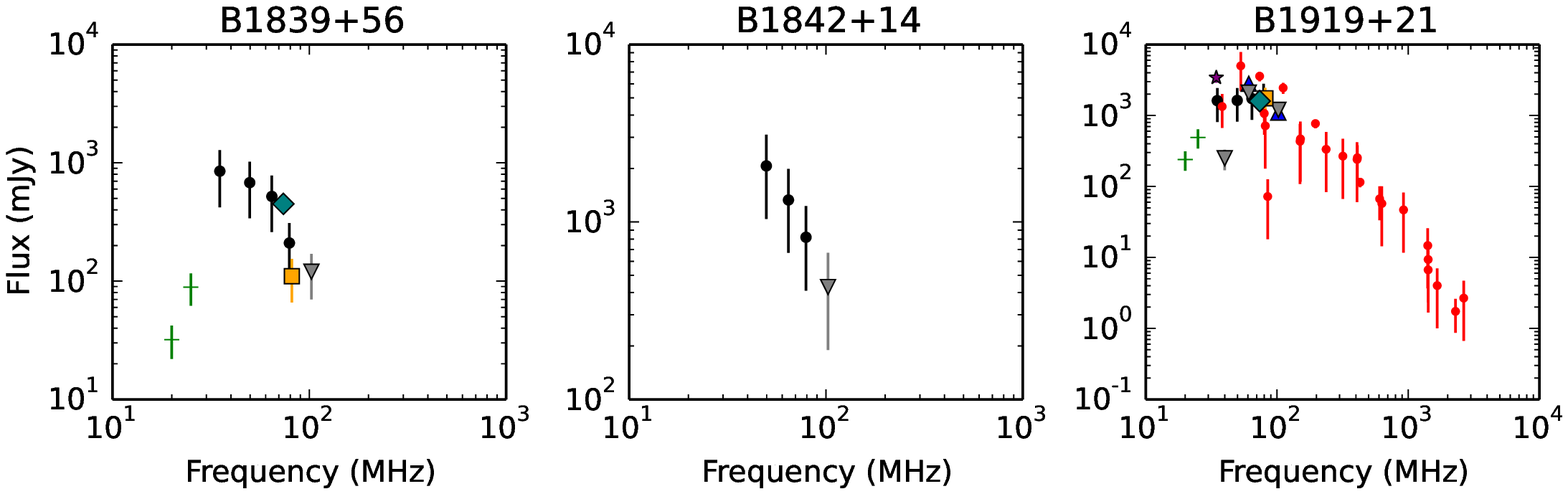}
        \includegraphics[width=0.9\textwidth]{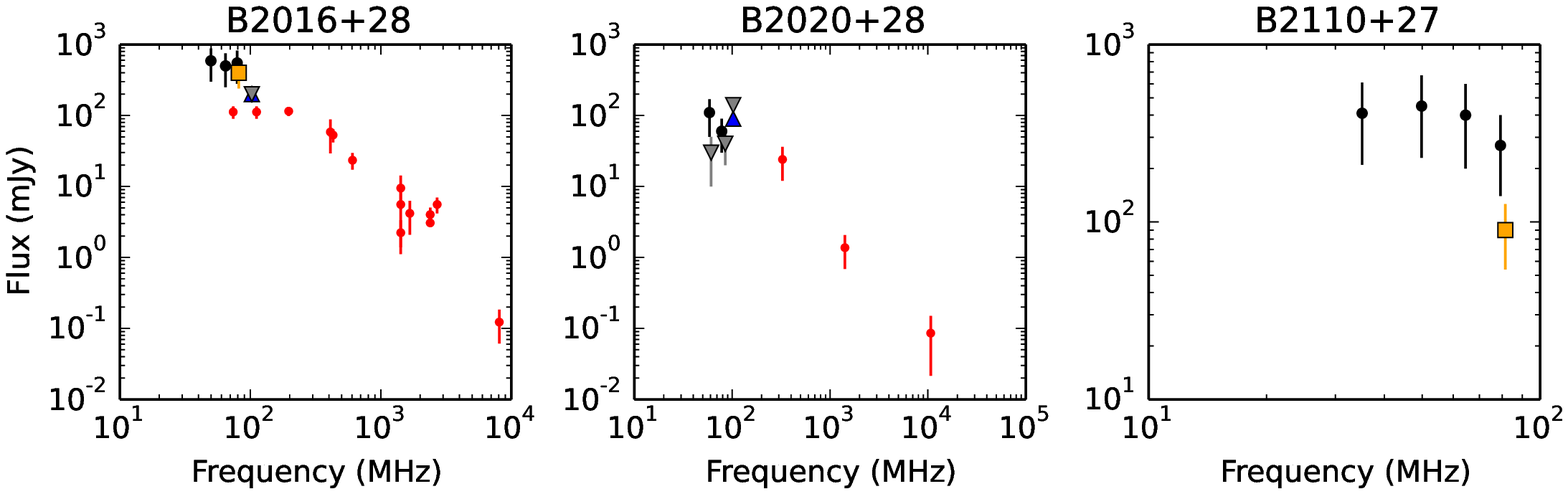}
        \includegraphics[width=0.9\textwidth]{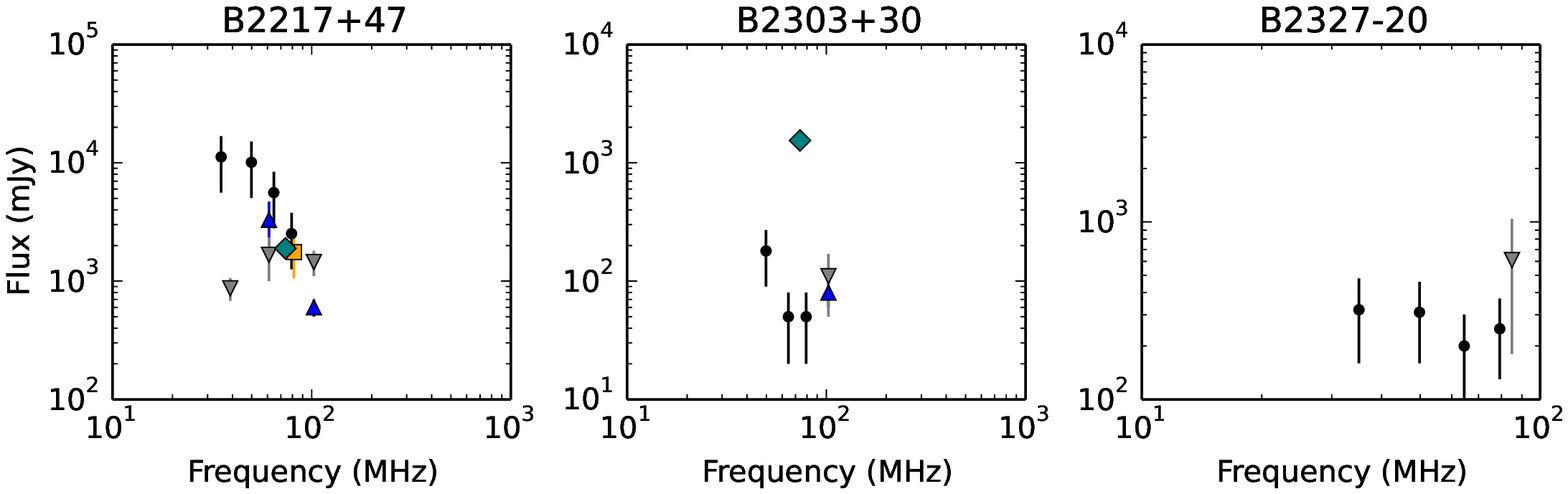}
\caption{Flux density measurements for 36 pulsars at a variety of frequencies obtained using LWA1 plotted are plotted with values at
comparable frequencies. Black filled circles are our measurements, red filled circles are from~\cite{1973A&A....28..237S}, blue up
triangles are from~\cite{1979SvA....23..179I}, gray down trianges are from~\cite{1981Ap&SS..78...45I},
purple stars are from~\cite{1992JApA...13..151D}, orange squares are from~\cite{1998ApJ...509..785S}, yellow pentagons are from~\cite{2011A&A...525A..55K}, green pluses are from~\cite{2013MNRAS.431.3624Z}, and teal diamonds are from~\cite{2014MNRAS.440..327L}.}
\label{fig:spectra1}
\end{figure*}
}

\clearpage

{\bfref For 27 of the pulsars, we measured mean flux densities at three or more frequencies and performed a linear
fit for a spectral index across all or a portion of the LWA1 band. The spectral indices that we determined are given
in Table~\ref{table:flux} and Figure~\ref{fig:histspec} shows a histogram of these values. The distribution of spectral
indices has a mean of $-$0.7 and a standard deviation of 1.0. A comparison of the mean of this low frequency distribution
to the mean ($-$1.41) of \cite{2013MNRAS.431.1352B} which was derived based on pulsar observations near 1 GHz provides further
evidence for turnover in pulsar spectra near 100 MHz. }

\begin{figure}[h!]
\centering
\includegraphics[width=0.5\textwidth]{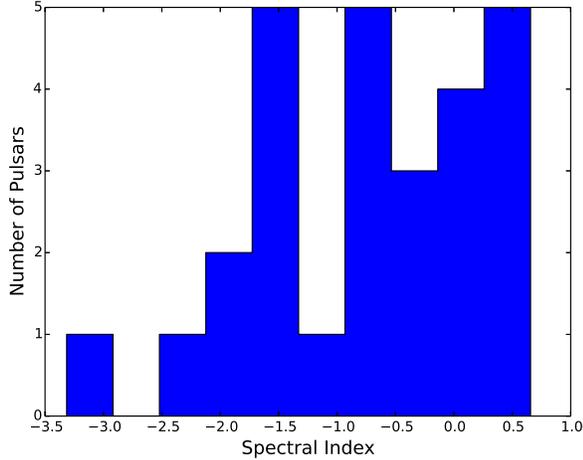}
\caption[Histogram of Spectral Indices]{The distribution of spectral indices in the frequency range 30 to 88 MHz for 27 pulsars.}
\label{fig:histspec}
\end{figure}
}

{\bfref \subsection{Pulse Component Analysis}\label{subsec:resultcomponents}
In Tables~\ref{table:widths2} and~\ref{table:widths3}, we present separation of components and
individual component widths for 2 and 3 component profiles, respectively. Ten of the detected profiles
required 2 Gaussian components while five required 3 Gaussian components. PSR B0943+10 is included in
both tables because in B mode, it has 3 components while in Q mode, only 2 components were required
during profile fitting. Combining the analysis of these individual components with the overall profiles
widths, $w_{50}$ and $w_{10}$, shows a general trend for these parameters to increase with decreasing
frequencies, supporting RFM. However, there are some exceptions, such as PSR B1919+21 which does not
show any increase in any of these parameters, but rather a decrease (with the exception of $w_{10}$)
as frequency decreases. This has been previously observed for PSR B1919+21~\cite[e.g.][]{2002ApJ...577..322M}.
Other exceptions include PSRs B0149$-$16 and B1112$+$50. Though the overall width of
PSR B0149$-$16 increases at lower frequency, the separation of its two components remains constant. The overall
increase is due to the increase in width of the components. In the single component PSR B1112$+$50, both
$w_{50}$ and $w_{10}$ are unchanged over the LWA1 frequency band.

In addition to evolution of the width of pulsar profiles as well as individual components, considerable
change in the relative amplitudes of individual components. In Table~\ref{table:widths2}, we compare
the relative amplitudes of the 2 components at various frequencies. PSRs B0031-07, B0809+74,
and B1919+21 all show considerable changes within the LWA1 observing band. Others, including B0149-16,
B0301+19 and B2327-20 seem to have a slight trend in profile evolution, while the other 4 pulsars listed
are relatively unchanged in the LWA1 band or have too few measurements reported here.

{ \footnotesize
\begin{longtable}{cccccc}
Pulsar & $\nu$ & $\Delta \phi_{12}$ & $w_{50}^{1}$ & $w_{50}^{2}$ & $A_{21}$\\
 & MHz & phase & phase & phase \\
\hline\hline
\endhead
B0031-07 & 35.1 & 0.111(7) & 0.08(1) & 0.070(5) & 2.2(5) \\
         & 49.8 & 0.10(1) & 0.11(2) & 0.068(5) & 1.4(2) \\
         & 64.5 & 0.085(6) & 0.079(8) & 0.063(6) & 1.3(2) \\
         & 79.2 & 0.07(1) & 0.047(7) & 0.09(1) & 1.3(3) \\
\hline
B0149-16 & 35.1 & 0.021(2) & 0.012(1) & 0.017(2) & 1.0(2) \\
         & 49.8 & 0.021(1) & 0.0094(6) & 0.018(2) & 0.8(1) \\
         & 64.5 & 0.0222(9) & 0.0082(5) & 0.013(2) & 0.6(1) \\
         & 79.2 & 0.020(1) & 0.0071(6) & 0.016(3) & 0.7(2) \\
\hline
B0301+19 & 49.8 & 0.062(4) & 0.030(6) & 0.027(4) & 1.2(4) \\
         & 64.5 & 0.058(2) & 0.013(2) & 0.011(2) & 0.9(3) \\
         & 79.2 & 0.053(2) & 0.014(2) & 0.008(2) & 0.6(2) \\
\hline
B0320+39 & 39.0 & 0.030(3) & 0.017(3) & 0.019(5) & 0.6(2)\\
         & 58.6 & 0.017(1) & 0.013(1) & 0.011(2) & 0.6(1)\\
         & 78.2 & 0.016(1) & 0.013(1) & 0.013(2) & 0.9(2)\\
\hline
B0525+21 & 39.0 & 0.063(2) & 0.030(3) & 0.029(3) & 0.9(1) \\
         & 58.6 & 0.0601(8) & 0.0153(7) & 0.019(1) & 0.88(8) \\
         & 68.0 & 0.054(2) & 0.012(1) & 0.012(2) & 0.6(2) \\
         & 78.2 & 0.0555(5) & 0.0107(5) & 0.0115(6) & 0.87(7) \\
\hline
B0809+74 & 35.0 & 0.077(5) & 0.069(9) & 0.056(3) & 1.8(3) \\
         & 45.0 & 0.068(4) & 0.048(5) & 0.054(4) & 1.7(3) \\
         & 55.0 & 0.055(2) & 0.036(1) & 0.060(2) & 2.0(2) \\
         & 75.0 & 0.027(2) & 0.026(2) & 0.066(1) & 4.8(8) \\
\hline
B0943+10Q & 35.1 & 0.035(2) & 0.032(3) & 0.032(1) & 2.4(4) \\
          & 64.5 & 0.026(3) & 0.029(3) & 0.031(1) & 3.2(9) \\
\hline
B1604-00 & 35.1 & 0.019(8) & 0.017(3) & 0.028(9) & 1.0(9) \\
         & 49.8 & 0.019(1) & 0.0142(7) & 0.017(1) & 0.7(1) \\
         & 64.5 & 0.0192(6) & 0.0101(5) & 0.016(1) & 1.0(1)\\
         & 79.2 & 0.0180(9) & 0.011(1) & 0.011(1) & 0.9(1) \\
\hline
B1919+21 & 55.0 & 0.0075(1) & 0.0061(1) & 0.0213(1) & 4.1(1) \\
         & 65.0 & 0.0081(1) & 0.00630(8) & 0.0201(1) & 3.5(1) \\
         & 75.0 & 0.0113(2) & 0.0085(1) & 0.0133(2) & 1.12(4) \\
\hline
B2327-20 & 35.1 & 0.012(2) & 0.010(2) & 0.006(1) & 0.5(2) \\
         & 49.8 & 0.010(2) & 0.006(2) & 0.009(4) & 0.6(3) \\
         & 64.5 & 0.0105(7) & 0.0061(8) & 0.0064(8) & 0.9(2) \\
         & 79.2 & 0.009(1) & 0.007(2) & 0.006(1) & 1.0(4) \\
\hline
\caption[Component spacing and widths.]{\bfref Component spacing and widths for 10 pulsars with two
components in the LWA1 frequency band.}
\label{table:widths2}
\end{longtable} }

\clearpage

{ \footnotesize
\begin{longtable*}{cccccccc}
Pulsar & $\nu$ & $\Delta \phi_{13}$ & $\Delta \phi_{12}$ & $\Delta \phi_{23}$ & $w_{50}^{1}$ & $w_{50}^{2}$ & $w_{50}^{3}$ \\
 & MHz & phase & phase & phase & phase & phase & phase \\
\hline\hline
\endhead
B0834+06 & 35.1 & 0.039(4) & 0.0156(4) & 0.024(4) & 0.0090(3) & 0.0218(7) & 0.084(7) \\
         & 49.8 & 0.020(1) & 0.0140(2) & 0.006(1) & 0.0086(2) & 0.0132(4) & 0.034(1) \\
         & 64.5 & 0.019(1) & 0.0127(1) & 0.006(1) & 0.0080(1) & 0.0121(3) & 0.031(1) \\
         & 79.2 & 0.020(4) & 0.0120(1) & 0.008(4) & 0.0078(1) & 0.0120(3) & 0.030(4) \\
\hline
B0943+10B & 35.0 & 0.0501(6) & 0.013(9) & 0.037(9) & 0.0254(8) & 0.13(2) & 0.024(1) \\
          & 45.0 & 0.0459(5) & 0.014(5) & 0.031(5) & 0.0214(5) & 0.08(1) & 0.0210(9) \\
          & 55.0 & 0.0396(5) & 0.017(4) & 0.023(4) & 0.0194(5) & 0.074(8) & 0.020(1) \\
          & 65.0 & 0.0369(4) & 0.022(4) & 0.015(4) & 0.0180(5) & 0.077(9) & 0.0178(9) \\
          & 75.0 & 0.0334(5) & 0.015(4) & 0.019(4) & 0.0169(7) & 0.08(1) & 0.018(1) \\
\hline
B0950+08 & 25.0 & 0.065(4) & 0.015(7) & 0.050(7) & 0.050(6) & 0.19(2) & 0.048(5) \\
         & 38.0 & 0.055(1) & -0.024(8) & 0.079(8) & 0.032(2) & 0.081(8) & 0.047(1) \\
         & 45.0 & 0.054(2) & -0.02(1) & 0.08(1) & 0.031(3) & 0.07(1) & 0.050(2) \\
         & 55.0 & 0.0512(9) & 0.007(3) & 0.045(3) & 0.031(2) & 0.130(5) & 0.030(1) \\
         & 65.0 & 0.043(2) & -0.02(1) & 0.07(1) & 0.025(2) & 0.07(1) & 0.037(2) \\
         & 74.0 & 0.0396(3) & 0.003(1) & 0.037(1) & 0.0214(5) & 0.109(2) & 0.0253(6) \\
\hline
B1133+16 & 35.1 & 0.0357(1) & 0.0259(9) & 0.0099(9) & 0.0127(3) & 0.063(2) & 0.0124(2) \\
         & 49.8 & 0.03317(7) & 0.0188(7) & 0.0143(6) & 0.0113(1) & 0.060(1) & 0.01138(7) \\
         & 64.5 & 0.03101(5) & 0.0151(7) & 0.0159(7) & 0.00943(8) & 0.047(1) & 0.01041(6) \\
         & 79.2 & 0.02917(5) & 0.0135(7) & 0.0157(7) & 0.00857(8) & 0.0412(9) & 0.00976(7) \\
\hline
B1237+25 & 35.1 & 0.0496(9) & 0.026(2) & 0.024(2) & 0.0097(7) & 0.009(4) & 0.011(1) \\
         & 49.8 & 0.0485(5) & 0.0270(5) & 0.0215(7) & 0.0094(3) & 0.0079(8) & 0.0127(8) \\
         & 64.5 & 0.0455(4) & 0.0261(4) & 0.0193(6) & 0.0084(2) & 0.0088(8) & 0.0109(7) \\
         & 79.2 & 0.0438(3) & 0.0249(6) & 0.0189(7) & 0.0074(2) & 0.012(1) & 0.0087(6) \\
\hline
\caption[Component spacing and widths.]{\bfref Component spacing and widths for 5 pulsars with 3 components in the LWA1 frequency range.}
\label{table:widths3}
\end{longtable*} }

}

\section{Conclusion}\label{sec:conclusion}
We have presented detections of {\bfref 44} pulsars using LWA1, including precise DM measurements for
all {\bfref 44} and mean flux density measurements at a variety of frequencies within 40--88 MHz for
{\bfref 36} of them. {\bfref As expected, our flux density measurements confirm spectral turnover in
many of the pulsars we have detected. The mean flux density measurements presented in this paper did not account
for the effect of LST on the LWA1 SEFD and therefore have larger errors than we believe to be possible
in the near future. We intend to enhance our SEFD model for LWA1, enabling much more precise measurements
of pulsar flux densities, which will allow improved studies of the spectral properties of pulsars.}
{\bfref We have also presented component spacing and widths for 15 profiles (14 pulsars
since B0943+10 has two modes) which required either 2 or 3 components to adequately model the profile. For
2-component profiles, we have also compared the ratio of the second peak to that of the first.} These
detections demonstrate the capabilities of LWA1 for use in pulsar astronomy work, including the detection
of 3 MSPs to date, J0030+0451, J0034-0534, and J2145-0750. LWA1's observing band goes down to the lowest
frequencies observable through the Earth's ionosphere. Observations at these frequencies are very important
for understanding the ISM as well as the pulsars themselves. A comparison of our measured DMs to those
reported in previous work show that the DM has changed significantly since the original measurement. The
derived rate of change is comparable to past DM variation measurements, however these changes can easily
be monitored {\bfref using} LWA1. {\bfref Further work needs to be done to account for possible bias in the
DM measurements due to profile evolution, however the observed difference in DM is too large to be due to
profile evolution alone. The observations presented here resulted in DM measurements with errors of order
$10^{-5} \mathrm{pc\;cm^{-3}}$, in some pulsars. Periodic measurements with this precision of many pulsars will enhance
our understanding of the ISM and may enable improved correction for DM variation of data sets used in attempts to
detect the signal of GWs using PTAs.} We have begun monitoring the 3 MSPs for DM variation over time and based
on the results {\bfref presented above}, we will also monitor pulsars that have shown significant changes and present those
results in future publications. We have {\bfref also} presented profiles for {\bfref 44} pulsars, some of which
show considerable
profile evolution over the LWA1 frequency band. Some of the profile evolution is likely due to interstellar
scattering which will be investigated further and presented in future work. Profile evolution seen
in PSRs B0031-07, B0320+39, and B0809+74 consists of considerable change in {\bfref the ratio of the amplitude
of the 2 components making up their profiles.} {\bfref An analysis of component width and spacing largely
supports RFM throughout the LWA1 band, however there are a few exceptions.} We have also introduced the open access LWA
Pulsar Data Archive, where we are archiving pulsar observations performed with LWA1 in a variety of
reduced data formats suitable for analysis of folded pulsar data as well as single pulse analysis.

\section*{Acknowledgements}
Construction of the LWA has been supported by the Office of Naval Research under Contract N00014-07-C-0147. Support for
operations and continuing development of the LWA1 is provided by the National Science Foundation under grants AST-1139963
and AST-1139974 of the University Radio Observatory program. Basic research on pulsars at NRL is supported by the Chief of
Naval Research (CNR). Part of this research was carried out at the Jet Propulsion Laboratory, California Institute of
Technology, under a contract with the National Aeronautics and Space Administration. The authors thank an anonymous referee
for their useful comments.

{\it Facilities:} \facility{LWA}

\bibliography{lwapulsars}

\begin{thebibliography}{56}
\providecommand\natexlab[1]{#1}
\providecommand\JournalTitle[1]{#1}

\bibitem[{{Ahuja} {et~al.}(2007){Ahuja}, {Mitra}, \&
  {Gupta}}]{2007MNRAS.377..677A}
{Ahuja}, A.~L., {Mitra}, D., \& {Gupta}, Y. 2007,
  \href{http://dx.doi.org/10.1111/j.1365-2966.2007.11630.x}{\JournalTitle{\mnras},
  377, 677}

\bibitem[{{Bates} {et~al.}(2013){Bates}, {Lorimer}, \&
  {Verbiest}}]{2013MNRAS.431.1352B}
{Bates}, S.~D., {Lorimer}, D.~R., \& {Verbiest}, J.~P.~W. 2013,
  \href{http://dx.doi.org/10.1093/mnras/stt257}{\JournalTitle{\mnras}, 431,
  1352}

\bibitem[{{Bhat} {et~al.}(2014){Bhat}, {Ord}, {Tremblay}, {Tingay},
  {Deshpande}, {van Straten}, {Oronsaye}, {Bernardi}, {Bowman}, {Briggs},
  {Cappallo}, {Corey}, {Emrich}, {Goeke}, {Greenhill}, {Hazelton}, {Hewitt},
  {Johnston-Hollitt}, {Kaplan}, {Kasper}, {Kratzenberg}, {Lonsdale}, {Lynch},
  {McWhirter}, {Mitchell}, {Morales}, {Morgan}, {Oberoi}, {Prabu}, {Rogers},
  {Roshi}, {Udaya Shankar}, {Srivani}, {Subrahmanyan}, {Waterson}, {Wayth},
  {Webster}, {Whitney}, {Williams}, \& {Williams}}]{2014ApJ...791L..32B}
{Bhat}, N.~D.~R., {Ord}, S.~M., {Tremblay}, S.~E., {et~al.} 2014,
  \href{http://dx.doi.org/10.1088/2041-8205/791/2/L32}{\JournalTitle{\apjl},
  791, L32}

\bibitem[{{Bilous} {et~al.}(2014){Bilous}, {Hessels}, {Kondratiev}, {van
  Leeuwen}, {Stappers}, {Weltevrede}, {Falcke}, {Hassall}, {Pilia}, {Keane},
  {Kramer}, {Grie{\ss}meier}, \& {Serylak}}]{2014A&A...572A..52B}
{Bilous}, A.~V., {Hessels}, J.~W.~T., {Kondratiev}, V.~I., {et~al.} 2014,
  \href{http://dx.doi.org/10.1051/0004-6361/201424425}{\JournalTitle{\aap},
  572, A52}

\bibitem[{{Braude} {et~al.}(1978){Braude}, {Megn}, {Riabov}, {Sharykin}, \&
  {Zhuk}}]{1978Ap&SS..54....3B}
{Braude}, S.~I., {Megn}, A.~V., {Riabov}, B.~P., {Sharykin}, N.~K., \& {Zhuk},
  I.~N. 1978, \href{http://dx.doi.org/10.1007/BF00637902}{\JournalTitle{\apss},
  54, 3}

\bibitem[{{Cordes}(1978)}]{1978ApJ...222.1006C}
{Cordes}, J.~M. 1978,
  \href{http://dx.doi.org/10.1086/156218}{\JournalTitle{\apj}, 222, 1006}

\bibitem[{{Cordes}(2002)}]{2002ASPC..278..227C}
{Cordes}, J.~M. 2002, in Astronomical Society of the Pacific Conference Series,
  Vol. 278, Single-Dish Radio Astronomy: Techniques and Applications, ed.
  S.~{Stanimirovic}, D.~{Altschuler}, P.~{Goldsmith}, \& C.~{Salter}, 227

\bibitem[{{Cordes} \& {Shannon}(2010)}]{2010arXiv1010.3785C}
{Cordes}, J.~M., \& {Shannon}, R.~M. 2010, \JournalTitle{ArXiv e-prints},
  \href{http://arxiv.org/abs/1010.3785}{{\sffamily arXiv:1010.3785
  [astro-ph.IM]}}

\bibitem[{{Cordes} {et~al.}(2015){Cordes}, {Shannon}, \&
  {Stinebring}}]{2015arXiv150308491C}
{Cordes}, J.~M., {Shannon}, R.~M., \& {Stinebring}, D.~R. 2015,
  \JournalTitle{ArXiv e-prints},
  \href{http://arxiv.org/abs/1503.08491}{{\sffamily arXiv:1503.08491
  [astro-ph.IM]}}

\bibitem[{{Dembska} {et~al.}(2014){Dembska}, {Kijak}, {Jessner}, {Lewandowski},
  {Bhattacharyya}, \& {Gupta}}]{2014MNRAS.445.3105D}
{Dembska}, M., {Kijak}, J., {Jessner}, A., {et~al.} 2014,
  \href{http://dx.doi.org/10.1093/mnras/stu1905}{\JournalTitle{\mnras}, 445,
  3105}

\bibitem[{{Demorest} {et~al.}(2013){Demorest}, {Ferdman}, {Gonzalez}, {Nice},
  {Ransom}, {Stairs}, {Arzoumanian}, {Brazier}, {Burke-Spolaor}, {Chamberlin},
  {Cordes}, {Ellis}, {Finn}, {Freire}, {Giampanis}, {Jenet}, {Kaspi}, {Lazio},
  {Lommen}, {McLaughlin}, {Palliyaguru}, {Perrodin}, {Shannon}, {Siemens},
  {Stinebring}, {Swiggum}, \& {Zhu}}]{2013ApJ...762...94D}
{Demorest}, P.~B., {Ferdman}, R.~D., {Gonzalez}, M.~E., {et~al.} 2013,
  \href{http://dx.doi.org/10.1088/0004-637X/762/2/94}{\JournalTitle{\apj}, 762,
  94}

\bibitem[{{Demorest}(2015)}]{2015Demorest}
{Demorest}, P.~B. e.~a. 2015, \JournalTitle{in prep.}

\bibitem[{{Deshpande} \& {Radhakrishnan}(1992)}]{1992JApA...13..151D}
{Deshpande}, A.~A., \& {Radhakrishnan}, V. 1992,
  \href{http://dx.doi.org/10.1007/BF02702307}{\JournalTitle{Journal of
  Astrophysics and Astronomy}, 13, 151}

\bibitem[{{Desvignes, G.}(2015)}]{2015Desvignes}
{Desvignes, G.}, e.~a. 2015, \JournalTitle{in prep.}

\bibitem[{{Dewey} {et~al.}(1985){Dewey}, {Taylor}, {Weisberg}, \&
  {Stokes}}]{1985ApJ...294L..25D}
{Dewey}, R.~J., {Taylor}, J.~H., {Weisberg}, J.~M., \& {Stokes}, G.~H. 1985,
  \href{http://dx.doi.org/10.1086/184502}{\JournalTitle{\apjl}, 294, L25}

\bibitem[{{Dowell} {et~al.}(2012){Dowell}, {Wood}, {Stovall}, {Ray}, {Clarke},
  \& {Taylor}}]{2012JAI.....150006D}
{Dowell}, J., {Wood}, D., {Stovall}, K., {et~al.} 2012,
  \href{http://dx.doi.org/10.1142/S2251171712500067}{\JournalTitle{Journal of
  Astronomical Instrumentation}, 1, 50006}

\bibitem[{{Dowell} {et~al.}(2013){Dowell}, {Ray}, {Taylor}, {Blythe}, {Clarke},
  {Craig}, {Ellingson}, {Helmboldt}, {Henning}, {Lazio}, {Schinzel}, {Stovall},
  \& {Wolfe}}]{2013ApJ...775L..28D}
{Dowell}, J., {Ray}, P.~S., {Taylor}, G.~B., {et~al.} 2013,
  \href{http://dx.doi.org/10.1088/2041-8205/775/1/L28}{\JournalTitle{\apjl},
  775, L28}

\bibitem[{{Ellingson} {et~al.}(2013{\natexlab{a}}){Ellingson}, {Clarke},
  {Craig}, {Hicks}, {Lazio}, {Taylor}, {Wilson}, \&
  {Wolfe}}]{2013ApJ...768..136E}
{Ellingson}, S.~W., {Clarke}, T.~E., {Craig}, J., {et~al.} 2013{\natexlab{a}},
  \href{http://dx.doi.org/10.1088/0004-637X/768/2/136}{\JournalTitle{\apj},
  768, 136}

\bibitem[{{Ellingson} {et~al.}(2013{\natexlab{b}}){Ellingson}, {Taylor},
  {Craig}, {Hartman}, {Dowell}, {Wolfe}, {Clarke}, {Hicks}, {Kassim}, {Ray},
  {Rickard}, {Schinzel}, \& {Weiler}}]{2013ITAP...61.2540E}
{Ellingson}, S.~W., {Taylor}, G.~B., {Craig}, J., {et~al.} 2013{\natexlab{b}},
  \href{http://dx.doi.org/10.1109/TAP.2013.2242826}{\JournalTitle{IEEE
  Transactions on Antennas and Propagation}, 61, 2540}

\bibitem[{Frigo \& Johnson(1998)}]{Frigo98fftw:an}
Frigo, M., \& Johnson, S.~G. 1998, in  (IEEE), 1381

\bibitem[{{Haslam} {et~al.}(1982){Haslam}, {Salter}, {Stoffel}, \&
  {Wilson}}]{1982A&AS...47....1H}
{Haslam}, C.~G.~T., {Salter}, C.~J., {Stoffel}, H., \& {Wilson}, W.~E. 1982,
  \JournalTitle{\aaps}, 47, 1

\bibitem[{{Hassall} {et~al.}(2012){Hassall}, {Stappers}, {Hessels}, {Kramer},
  {Alexov}, {Anderson}, {Coenen}, {Karastergiou}, {Keane}, {Kondratiev},
  {Lazaridis}, {van Leeuwen}, {Noutsos}, {Serylak}, {Sobey}, {Verbiest},
  {Weltevrede}, {Zagkouris}, {Fender}, {Wijers}, {B{\"a}hren}, {Bell},
  {Broderick}, {Corbel}, {Daw}, {Dhillon}, {Eisl{\"o}ffel}, {Falcke},
  {Grie{\ss}meier}, {Jonker}, {Law}, {Markoff}, {Miller-Jones}, {Osten}, {Rol},
  {Scaife}, {Scheers}, {Schellart}, {Spreeuw}, {Swinbank}, {ter Veen}, {Wise},
  {Wijnands}, {Wucknitz}, {Zarka}, {Asgekar}, {Bell}, {Bentum}, {Bernardi},
  {Best}, {Bonafede}, {Boonstra}, {Brentjens}, {Brouw}, {Br{\"u}ggen},
  {Butcher}, {Ciardi}, {Garrett}, {Gerbers}, {Gunst}, {van Haarlem}, {Heald},
  {Hoeft}, {Holties}, {de Jong}, {Koopmans}, {Kuniyoshi}, {Kuper}, {Loose},
  {Maat}, {Masters}, {McKean}, {Meulman}, {Mevius}, {Munk}, {Noordam},
  {Orr{\'u}}, {Paas}, {Pandey-Pommier}, {Pandey}, {Pizzo}, {Polatidis},
  {Reich}, {R{\"o}ttgering}, {Sluman}, {Steinmetz}, {Sterks}, {Tagger}, {Tang},
  {Tasse}, {Vermeulen}, {van Weeren}, {Wijnholds}, \&
  {Yatawatta}}]{2012A&A...543A..66H}
{Hassall}, T.~E., {Stappers}, B.~W., {Hessels}, J.~W.~T., {et~al.} 2012,
  \href{http://dx.doi.org/10.1051/0004-6361/201218970}{\JournalTitle{\aap},
  543, A66}

\bibitem[{{Hewish} {et~al.}(1968){Hewish}, {Bell}, {Pilkington}, {Scott}, \&
  {Collins}}]{1968Natur.217..709H}
{Hewish}, A., {Bell}, S.~J., {Pilkington}, J.~D.~H., {Scott}, P.~F., \&
  {Collins}, R.~A. 1968,
  \href{http://dx.doi.org/10.1038/217709a0}{\JournalTitle{\nat}, 217, 709}

\bibitem[{{Hotan} {et~al.}(2004){Hotan}, {van Straten}, \&
  {Manchester}}]{2004PASA...21..302H}
{Hotan}, A.~W., {van Straten}, W., \& {Manchester}, R.~N. 2004,
  \href{http://dx.doi.org/10.1071/AS04022}{\JournalTitle{\pasa}, 21, 302}

\bibitem[{{Izvekova} {et~al.}(1981){Izvekova}, {Kuzmin}, {Malofeev}, \&
  {Shitov}}]{1981Ap&SS..78...45I}
{Izvekova}, V.~A., {Kuzmin}, A.~D., {Malofeev}, V.~M., \& {Shitov}, I.~P. 1981,
  \href{http://dx.doi.org/10.1007/BF00654022}{\JournalTitle{\apss}, 78, 45}

\bibitem[{{Izvekova} {et~al.}(1979){Izvekova}, {Kuz'min}, {Malofeev}, \&
  {Shitov}}]{1979SvA....23..179I}
{Izvekova}, V.~A., {Kuz'min}, A.~D., {Malofeev}, V.~M., \& {Shitov}, Y.~P.
  1979, \JournalTitle{\sovast}, 23, 179

\bibitem[{{Karastergiou} \& {Johnston}(2007)}]{2007MNRAS.380.1678K}
{Karastergiou}, A., \& {Johnston}, S. 2007,
  \href{http://dx.doi.org/10.1111/j.1365-2966.2007.12237.x}{\JournalTitle{\mnras},
  380, 1678}

\bibitem[{{Karuppusamy} {et~al.}(2011){Karuppusamy}, {Stappers}, \&
  {Serylak}}]{2011A&A...525A..55K}
{Karuppusamy}, R., {Stappers}, B.~W., \& {Serylak}, M. 2011,
  \href{http://dx.doi.org/10.1051/0004-6361/201014507}{\JournalTitle{\aap},
  525, A55}

\bibitem[{{Kijak} {et~al.}(2007){Kijak}, {Gupta}, \&
  {Krzeszowski}}]{2007A&A...462..699K}
{Kijak}, J., {Gupta}, Y., \& {Krzeszowski}, K. 2007,
  \href{http://dx.doi.org/10.1051/0004-6361:20066125}{\JournalTitle{\aap}, 462,
  699}

\bibitem[{{Lane} {et~al.}(2014){Lane}, {Cotton}, {van Velzen}, {Clarke},
  {Kassim}, {Helmboldt}, {Lazio}, \& {Cohen}}]{2014MNRAS.440..327L}
{Lane}, W.~M., {Cotton}, W.~D., {van Velzen}, S., {et~al.} 2014,
  \href{http://dx.doi.org/10.1093/mnras/stu256}{\JournalTitle{\mnras}, 440,
  327}

\bibitem[{{L{\"o}hmer} {et~al.}(2001){L{\"o}hmer}, {Kramer}, {Mitra},
  {Lorimer}, \& {Lyne}}]{2001ApJ...562L.157L}
{L{\"o}hmer}, O., {Kramer}, M., {Mitra}, D., {Lorimer}, D.~R., \& {Lyne}, A.~G.
  2001, \href{http://dx.doi.org/10.1086/338324}{\JournalTitle{\apjl}, 562,
  L157}

\bibitem[{{Lyne} \& {Manchester}(1988)}]{1988MNRAS.234..477L}
{Lyne}, A.~G., \& {Manchester}, R.~N. 1988, \JournalTitle{\mnras}, 234, 477

\bibitem[{{Malofeev} {et~al.}(1994){Malofeev}, {Gil}, {Jessner}, {Malov},
  {Seiradakis}, {Sieber}, \& {Wielebinski}}]{1994A&A...285..201M}
{Malofeev}, V.~M., {Gil}, J.~A., {Jessner}, A., {et~al.} 1994,
  \JournalTitle{\aap}, 285, 201

\bibitem[{{Manchester} {et~al.}(2012){Manchester}, {Hobbs}, {Bailes}, {Coles},
  {van Straten}, {Keith}, {Shannon}, {Bhat}, {Brown}, {Burke-Spolaor},
  {Champion}, {Chaudhary}, {Edwards}, {Hampson}, {Hotan}, {Jameson}, {Jenet},
  {Kesteven}, {Khoo}, {Kocz}, {Maciesiak}, {Oslowski}, {Ravi}, {Reynolds},
  {Sarkissian}, {Verbiest}, {Wen}, {Wilson}, {Yardley}, {Yan}, \&
  {You}}]{2012arXiv1210.6130M}
{Manchester}, R.~N., {Hobbs}, G., {Bailes}, M., {et~al.} 2012,
  \JournalTitle{ArXiv e-prints},
  \href{http://arxiv.org/abs/1210.6130}{{\sffamily arXiv:1210.6130
  [astro-ph.IM]}}

\bibitem[{{Mitra} \& {Rankin}(2002)}]{2002ApJ...577..322M}
{Mitra}, D., \& {Rankin}, J.~M. 2002,
  \href{http://dx.doi.org/10.1086/342136}{\JournalTitle{\apj}, 577, 322}

\bibitem[{{Phillips}(1991)}]{1991ApJ...373L..63P}
{Phillips}, J.~A. 1991,
  \href{http://dx.doi.org/10.1086/186052}{\JournalTitle{\apjl}, 373, L63}

\bibitem[{{Phillips} \& {Wolszczan}(1991)}]{1991ApJ...382L..27P}
{Phillips}, J.~A., \& {Wolszczan}, A. 1991,
  \href{http://dx.doi.org/10.1086/186206}{\JournalTitle{\apjl}, 382, L27}

\bibitem[{{Pilia} {et~al.}(2015){Pilia}, {Hessels}, {Stappers}, {Kondratiev},
  {Kramer}, {van~Leeuwen}, \& {Lyne}}]{2015Pilia}
{Pilia}, M., {Hessels}, J.~W.~T., {Stappers}, B.~W., {et~al.} 2015,
  \JournalTitle{A \& A, submitted}

\bibitem[{{Rankin}(1983)}]{1983ApJ...274..333R}
{Rankin}, J.~M. 1983,
  \href{http://dx.doi.org/10.1086/161450}{\JournalTitle{\apj}, 274, 333}

\bibitem[{{Ransom}(2001)}]{2001PhDT.......123R}
{Ransom}, S.~M. 2001, PhD thesis, Harvard University

\bibitem[{{Ryabov} {et~al.}(2010){Ryabov}, {Vavriv}, {Zarka}, {Ryabov},
  {Kozhin}, {Vinogradov}, \& {Denis}}]{2010A&A...510A..16R}
{Ryabov}, V.~B., {Vavriv}, D.~M., {Zarka}, P., {et~al.} 2010,
  \href{http://dx.doi.org/10.1051/0004-6361/200913335}{\JournalTitle{\aap},
  510, A16}

\bibitem[{{Schinzel} \& {Dowell}(2014)}]{memo201}
{Schinzel}, F., \& {Dowell}, J. 2014, \JournalTitle{LWA Memo Series}

\bibitem[{{Schinzel} \& {Polisensky}(2014)}]{memo202}
{Schinzel}, F., \& {Polisensky}, E. 2014, \JournalTitle{LWA Memo Series}

\bibitem[{{Shrauner} {et~al.}(1998){Shrauner}, {Taylor}, \&
  {Woan}}]{1998ApJ...509..785S}
{Shrauner}, J.~A., {Taylor}, J.~H., \& {Woan}, G. 1998,
  \href{http://dx.doi.org/10.1086/306510}{\JournalTitle{\apj}, 509, 785}

\bibitem[{{Sieber}(1973)}]{1973A&A....28..237S}
{Sieber}, W. 1973, \JournalTitle{\aap}, 28, 237

\bibitem[{{Stappers} {et~al.}(2011){Stappers}, {Hessels}, {Alexov}, {Anderson},
  {Coenen}, {Hassall}, {Karastergiou}, {Kondratiev}, {Kramer}, {van Leeuwen},
  {Mol}, {Noutsos}, {Romein}, {Weltevrede}, {Fender}, {Wijers}, {B{\"a}hren},
  {Bell}, {Broderick}, {Daw}, {Dhillon}, {Eisl{\"o}ffel}, {Falcke},
  {Griessmeier}, {Law}, {Markoff}, {Miller-Jones}, {Scheers}, {Spreeuw},
  {Swinbank}, {Ter Veen}, {Wise}, {Wucknitz}, {Zarka}, {Anderson}, {Asgekar},
  {Avruch}, {Beck}, {Bennema}, {Bentum}, {Best}, {Bregman}, {Brentjens}, {van
  de Brink}, {Broekema}, {Brouw}, {Br{\"u}ggen}, {de Bruyn}, {Butcher},
  {Ciardi}, {Conway}, {Dettmar}, {van Duin}, {van Enst}, {Garrett}, {Gerbers},
  {Grit}, {Gunst}, {van Haarlem}, {Hamaker}, {Heald}, {Hoeft}, {Holties},
  {Horneffer}, {Koopmans}, {Kuper}, {Loose}, {Maat}, {McKay-Bukowski},
  {McKean}, {Miley}, {Morganti}, {Nijboer}, {Noordam}, {Norden}, {Olofsson},
  {Pandey-Pommier}, {Polatidis}, {Reich}, {R{\"o}ttgering}, {Schoenmakers},
  {Sluman}, {Smirnov}, {Steinmetz}, {Sterks}, {Tagger}, {Tang}, {Vermeulen},
  {Vermaas}, {Vogt}, {de Vos}, {Wijnholds}, {Yatawatta}, \&
  {Zensus}}]{2011A&A...530A..80S}
{Stappers}, B.~W., {Hessels}, J.~W.~T., {Alexov}, A., {et~al.} 2011,
  \href{http://dx.doi.org/10.1051/0004-6361/201116681}{\JournalTitle{\aap},
  530, A80}

\bibitem[{{Suleymanova} \& {Izvekova}(1984)}]{SI84}
{Suleymanova}, S.~A., \& {Izvekova}, V.~A. 1984, \JournalTitle{Sov. Astron.},
  28, 53

\bibitem[{{Taylor} {et~al.}(2012){Taylor}, {Ellingson}, {Kassim}, {Craig},
  {Dowell}, {Wolfe}, {Hartman}, {Bernardi}, {Clarke}, {Cohen}, {Dalal},
  {Erickson}, {Hicks}, {Greenhill}, {Jacoby}, {Lane}, {Lazio}, {Mitchell},
  {Navarro}, {Ord}, {Pihlstr{\"o}m}, {Polisensky}, {Ray}, {Rickard},
  {Schinzel}, {Schmitt}, {Sigman}, {Soriano}, {Stewart}, {Stovall}, {Tremblay},
  {Wang}, {Weiler}, {White}, \& {Wood}}]{2012JAI.....150004T}
{Taylor}, G.~B., {Ellingson}, S.~W., {Kassim}, N.~E., {et~al.} 2012,
  \href{http://dx.doi.org/10.1142/S2251171712500043}{\JournalTitle{Journal of
  Astronomical Instrumentation}, 1, 50004}

\bibitem[{{Taylor}(1992)}]{1992PTRSL.341..117T}
{Taylor}, J.~H. 1992, \JournalTitle{Philosophical Transactions of the Royal
  Society of London, 341, 117-134 (1992)}, 341, 117

\bibitem[{{Thorsett}(1991)}]{1991ApJ...377..263T}
{Thorsett}, S.~E. 1991,
  \href{http://dx.doi.org/10.1086/170355}{\JournalTitle{\apj}, 377, 263}

\bibitem[{{Tingay} {et~al.}(2013){Tingay}, {Goeke}, {Bowman}, {Emrich}, {Ord},
  {Mitchell}, {Morales}, {Booler}, {Crosse}, {Wayth}, {Lonsdale}, {Tremblay},
  {Pallot}, {Colegate}, {Wicenec}, {Kudryavtseva}, {Arcus}, {Barnes},
  {Bernardi}, {Briggs}, {Burns}, {Bunton}, {Cappallo}, {Corey}, {Deshpande},
  {Desouza}, {Gaensler}, {Greenhill}, {Hall}, {Hazelton}, {Herne}, {Hewitt},
  {Johnston-Hollitt}, {Kaplan}, {Kasper}, {Kincaid}, {Koenig}, {Kratzenberg},
  {Lynch}, {Mckinley}, {Mcwhirter}, {Morgan}, {Oberoi}, {Pathikulangara},
  {Prabu}, {Remillard}, {Rogers}, {Roshi}, {Salah}, {Sault}, {Udaya-Shankar},
  {Schlagenhaufer}, {Srivani}, {Stevens}, {Subrahmanyan}, {Waterson},
  {Webster}, {Whitney}, {Williams}, {Williams}, \&
  {Wyithe}}]{2013PASA...30....7T}
{Tingay}, S.~J., {Goeke}, R., {Bowman}, J.~D., {et~al.} 2013,
  \href{http://dx.doi.org/10.1017/pasa.2012.007}{\JournalTitle{\pasa}, 30, 7}

\bibitem[{{Tremblay} {et~al.}(2015){Tremblay}, {Ord}, {Bhat}, {Tingay},
  {Crosse}, {Pallot}, {Oronsaye}, {Bernardi}, {Bowman}, {Briggs}, {Cappallo},
  {Corey}, {Deshpande}, {Emrich}, {Goeke}, {Greenhill}, {Hazelton},
  {Johnston-Hollitt}, {Kaplan}, {Kasper}, {Kratzenberg}, {Lonsdale}, {Lynch},
  {McWhirter}, {Mitchell}, {Morales}, {Morgan}, {Oberoi}, {Prabu}, {Rogers},
  {Roshi}, {Udaya Shankar}, {Srivani}, {Subrahmanyan}, {Waterson}, {Wayth},
  {Webster}, {Whitney}, {Williams}, \& {Williams}}]{2015PASA...32....5T}
{Tremblay}, S.~E., {Ord}, S.~M., {Bhat}, N.~D.~R., {et~al.} 2015,
  \href{http://dx.doi.org/10.1017/pasa.2015.6}{\JournalTitle{\pasa}, 32, 5}

\bibitem[{{Tsai} {et~al.}(2015){Tsai}, {Simonetti}, {Akukwe}, {Bear},
  {Cutchin}, {Dowell}, {Gough}, {Kanner}, {Kassim}, {Schinzel}, {Shawhan},
  {Taylor}, {Yancey}, {Quezada}, \& {Kavic}}]{2015AJ....149...65T}
{Tsai}, J.-W., {Simonetti}, J.~H., {Akukwe}, B., {et~al.} 2015,
  \href{http://dx.doi.org/10.1088/0004-6256/149/2/65}{\JournalTitle{\aj}, 149,
  65}

\bibitem[{{van Haarlem} {et~al.}(2013){van Haarlem}, {Wise}, {Gunst}, {Heald},
  {McKean}, {Hessels}, {de Bruyn}, {Nijboer}, {Swinbank}, {Fallows},
  {Brentjens}, {Nelles}, {Beck}, {Falcke}, {Fender}, {H{\"o}randel},
  {Koopmans}, {Mann}, {Miley}, {R{\"o}ttgering}, {Stappers}, {Wijers},
  {Zaroubi}, {van den Akker}, {Alexov}, {Anderson}, {Anderson}, {van Ardenne},
  {Arts}, {Asgekar}, {Avruch}, {Batejat}, {B{\"a}hren}, {Bell}, {Bell}, {van
  Bemmel}, {Bennema}, {Bentum}, {Bernardi}, {Best}, {B{\^i}rzan}, {Bonafede},
  {Boonstra}, {Braun}, {Bregman}, {Breitling}, {van de Brink}, {Broderick},
  {Broekema}, {Brouw}, {Br{\"u}ggen}, {Butcher}, {van Cappellen}, {Ciardi},
  {Coenen}, {Conway}, {Coolen}, {Corstanje}, {Damstra}, {Davies}, {Deller},
  {Dettmar}, {van Diepen}, {Dijkstra}, {Donker}, {Doorduin}, {Dromer}, {Drost},
  {van Duin}, {Eisl{\"o}ffel}, {van Enst}, {Ferrari}, {Frieswijk}, {Gankema},
  {Garrett}, {de Gasperin}, {Gerbers}, {de Geus}, {Grie{\ss}meier}, {Grit},
  {Gruppen}, {Hamaker}, {Hassall}, {Hoeft}, {Holties}, {Horneffer}, {van der
  Horst}, {van Houwelingen}, {Huijgen}, {Iacobelli}, {Intema}, {Jackson},
  {Jelic}, {de Jong}, {Juette}, {Kant}, {Karastergiou}, {Koers}, {Kollen},
  {Kondratiev}, {Kooistra}, {Koopman}, {Koster}, {Kuniyoshi}, {Kramer},
  {Kuper}, {Lambropoulos}, {Law}, {van Leeuwen}, {Lemaitre}, {Loose}, {Maat},
  {Macario}, {Markoff}, {Masters}, {McFadden}, {McKay-Bukowski}, {Meijering},
  {Meulman}, {Mevius}, {Middelberg}, {Millenaar}, {Miller-Jones}, {Mohan},
  {Mol}, {Morawietz}, {Morganti}, {Mulcahy}, {Mulder}, {Munk}, {Nieuwenhuis},
  {van Nieuwpoort}, {Noordam}, {Norden}, {Noutsos}, {Offringa}, {Olofsson},
  {Omar}, {Orr{\'u}}, {Overeem}, {Paas}, {Pandey-Pommier}, {Pandey}, {Pizzo},
  {Polatidis}, {Rafferty}, {Rawlings}, {Reich}, {de Reijer}, {Reitsma},
  {Renting}, {Riemers}, {Rol}, {Romein}, {Roosjen}, {Ruiter}, {Scaife}, {van
  der Schaaf}, {Scheers}, {Schellart}, {Schoenmakers}, {Schoonderbeek},
  {Serylak}, {Shulevski}, {Sluman}, {Smirnov}, {Sobey}, {Spreeuw}, {Steinmetz},
  {Sterks}, {Stiepel}, {Stuurwold}, {Tagger}, {Tang}, {Tasse}, {Thomas},
  {Thoudam}, {Toribio}, {van der Tol}, {Usov}, {van Veelen}, {van der Veen},
  {ter Veen}, {Verbiest}, {Vermeulen}, {Vermaas}, {Vocks}, {Vogt}, {de Vos},
  {van der Wal}, {van Weeren}, {Weggemans}, {Weltevrede}, {White}, {Wijnholds},
  {Wilhelmsson}, {Wucknitz}, {Yatawatta}, {Zarka}, {Zensus}, \& {van
  Zwieten}}]{2013A&A...556A...2V}
{van Haarlem}, M.~P., {Wise}, M.~W., {Gunst}, A.~W., {et~al.} 2013,
  \href{http://dx.doi.org/10.1051/0004-6361/201220873}{\JournalTitle{\aap},
  556, A2}

\bibitem[{{van Haasteren} {et~al.}(2011){van Haasteren}, {Levin}, {Janssen},
  {Lazaridis}, {Kramer}, {Stappers}, {Desvignes}, {Purver}, {Lyne}, {Ferdman},
  {Jessner}, {Cognard}, {Theureau}, {D'Amico}, {Possenti}, {Burgay},
  {Corongiu}, {Hessels}, {Smits}, \& {Verbiest}}]{2011MNRAS.414.3117V}
{van Haasteren}, R., {Levin}, Y., {Janssen}, G.~H., {et~al.} 2011,
  \href{http://dx.doi.org/10.1111/j.1365-2966.2011.18613.x}{\JournalTitle{\mnras},
  414, 3117}

\bibitem[{{Zakharenko} {et~al.}(2013){Zakharenko}, {Vasylieva}, {Konovalenko},
  {Ulyanov}, {Serylak}, {Zarka}, {Grie{\ss}meier}, {Cognard}, \&
  {Nikolaenko}}]{2013MNRAS.431.3624Z}
{Zakharenko}, V.~V., {Vasylieva}, I.~Y., {Konovalenko}, A.~A., {et~al.} 2013,
  \href{http://dx.doi.org/10.1093/mnras/stt470}{\JournalTitle{\mnras}, 431,
  3624}

\end{thebibliography}
\bibliographystyle{yahapj}

\end{document}